\documentclass[12pt]{article}
\usepackage{amsmath, amssymb, graphics}
\usepackage{amsbsy}
\usepackage{amsfonts}
\usepackage{amsmath}
\usepackage{graphicx}
\usepackage{array}
\usepackage{booktabs}
\usepackage{float}
\newcommand{\mathsym}[1]{{}}

\renewcommand{\title}[1]{\vbox{\center\LARGE{#1}}\vspace{5mm}}
\renewcommand{\author}[1]{\vbox{\center#1}\vspace{5mm}}

\makeatletter
\renewcommand\section{\@startsection {section}{1}{\z@}%
                                   {-3.5ex \@plus -1ex \@minus -.2ex}%
                                   {2.3ex \@plus.2ex}%
                                   {\normalfont\large\bfseries}}
\renewcommand\subsection{\@startsection{subsection}{2}{\z@}%
                                   {-3.25ex\@plus -1ex \@minus -.2ex}%
                                   {1.5ex \@plus .2ex}%
                                   {\normalfont\normalsize\bfseries}}
\makeatother

\renewcommand{\theequation}{1.\arabic{equation}}

\renewcommand{\[}{\begin{eqnarray}}
\renewcommand{\]}{\end{eqnarray}}
\newcommand{\nn}{\nonumber}

\newcommand{\cN}{\mathcal{N}}

\def\moth{\mathsurround=0pt}
\newdimen\zo \zo=0pt

\def\tick{\leaders\hrule height 0.5ex depth 0pt \hskip 0.5pt}
\def\upboxfill{$\moth \setbox\zo\hbox{\tick}%
  \hskip 2pt\hbox to 0pt{$\tick$\hss}\hrulefill \hbox to 2pt{$\tick$\hss}$}

\def\dtick{\leaders\hrule height .34pt depth 0.5ex \hskip 0.5pt}
\def\downboxfill{$\moth \setbox\zo\hbox{\dtick}%
  \hskip 2pt\hbox to 0pt{$\dtick$\hss}\hrulefill%
  \hbox to 2pt{$\dtick$\hss}$}

\newcommand{\bea}{\begin{eqnarray}}
\newcommand{\eea}{\end{eqnarray}}
\newcommand{\ee}{\end{equation}}
\newcommand{\be}{\begin{equation}}

\newcommand{\no}{\nonumber}

\def\cN{{\cal N}}

\def\a{\alpha}

\def\l{\lambda}

\def\no{\nonumber}

\def\Tr{{\rm Tr}}

\def\NeqFour{{{\cal N}=4}}

\def\NeqEight{{{\cal N}=8}}
\def\NeqTwo{{{\cal N}=2}}

\def\NeqOne{{{\cal N}=1}}

\def\flag{{f}}

\def\Ansatz{ansatz}

\def\tabentry#1#2#3{$\begin{array}{c}\text{#1} \\[-#3pt] \text{#2}\end{array}$}
\def\tabentry#1#2{$\begin{array}{c}\text{#1} \\[-5pt] \text{#2}\end{array}$}

\begin{document}

\textwidth 170mm
\textheight 230mm
\topmargin -1cm
\oddsidemargin-0.8cm \evensidemargin -0.8cm
\topskip 9mm
\headsep9pt

\overfullrule=0pt
\parskip=2pt
\parindent=12pt
\headheight=0in \headsep=0in \topmargin=0in \oddsidemargin=0in

\vspace{ -3cm} \thispagestyle{empty} \vspace{-1cm}

\begin{flushright}
IGC-12/11-5
\hfill
SU-ITP-12/41
\end{flushright}

\medskip

\begin{center}

{\Large \bf 
One-loop four-point amplitudes in pure and matter-coupled $\cN \leq 4$ supergravity}

\medskip

\vskip .2in

{\large John Joseph M. Carrasco$^{a}$, Marco Chiodaroli$^{b}$, Murat G\"{u}naydin$^b$ \\ and Radu Roiban$^b$}

\bigskip
\bigskip

$^{a}${ 
\small
Stanford Institute for Theoretical Physics and Department of Physics,\\
Stanford University, Stanford, CA 94305, USA \\
\texttt{jjmc@stanford.edu};}\\

\bigskip

$^{b}${ \small  Institute for Gravitation and the Cosmos \\
 The Pennsylvania State University, University Park PA 16802, USA \\ 
{\texttt{ mchiodar,\,murat,\,radu@phys.psu.edu};}
}\\

\end{center}

\bigskip


\begin{abstract}

We construct {\em all} supergravity theories that can be obtained through factorized orbifold projections 
of $\NeqEight$ supergravity, exposing their double-copy structure, and calculate their one-loop 
four-point scattering amplitudes.  We observe a unified structure in both matter and gravity amplitudes,
and demonstrate that the four-graviton amplitudes are insensitive to the
precise nature of the matter couplings.   We  show that these amplitudes are identical for the two 
different realizations of $\NeqFour$ supergravity with two vector multiplets, and argue that 
this feature extends to all multiplicities and loop orders as well as to higher dimensions.  We 
also construct a selected set of supergravities obtained through a non-factorized orbifold action.
Furthermore we calculate one-loop four-point amplitudes for all pure super-Yang-Mills theories with 
less-than-maximal supersymmetry using the duality between color and kinematics,  finding here a unified expression that holds for all four gluon amplitudes in these theories.
We  recover the related amplitudes of factorized ${\cal N}\leq4$ 
supergravities employing the double-copy construction.  We observe a requirement that the four-point loop-level  
amplitudes have non-local integrand representations, exhibiting a mild non-locality 
in the form of inverse powers of the three external Mandelstam invariants. 
These are the first  loop-level color-kinematic-satisfying representations in reduced supersymmetry theories.

\end{abstract}

\baselineskip=16pt
\setcounter{equation}{0}
\setcounter{footnote}{0}

\newpage

\renewcommand{\theequation}{1.\arabic{equation}}
\setcounter{equation}{0}

\section{Introduction}

This last decade has marked tremendous progress in multi-loop scattering amplitude 
calculations, especially for maximally supersymmetric gauge and gravitational theories.  This has  been 
accompanied by  the discovery of  novel mathematical structures with potentially broad applicability such as  integrability in the planar
$\cN =4$ super Yang-Mills (sYM) theory~\cite{integrability}, on-shell recursion~\cite{OSR},
dual superconformal and Yangian symmetries~\cite{Yangian},  a Grassmannian description of the S-matrix~\cite{grassmanian}, and the 
duality between color and kinematics~\cite{BCJ}.  Most notably, recent calculations have revealed a better-than-expected 
ultraviolet (UV) behavior for amplitudes in $\cN=8$ supergravity up to at least four loops \cite{gravity3, gravity3m, gravity4, gravitysum}
which has in turn lead to the conjecture that $\cN=8$ supergravity may be perturbatively 
finite \cite{BDR}\footnote{ Subsequent duality arguments have also been used to show that candidate counterterms for $\cN=8$ supergravity cannot appear 
until seven loops \cite{Kallosh:2009db,Green:2010sp,Green:2010wi,Bossard:2011ij}, as well as to 
argue in favor of the conjectured perturbative finiteness of the theory \cite{DualityArguments,Kallosh:2011dp,Kallosh:2011qt}, 
although subtleties with the decoupling of massive string states may alter this conclusion \cite{GSO}.}.  

As new techniques to study amplitudes in maximally  supersymmetric theories are developed, it
is natural to ask whether they can lead to new insight and to the discovery of novel structures in  theories with reduced supersymmetry. 
At the same time, the conjectured finiteness of $\cN=8$ supergravity 
raises the question of whether more phenomenologically viable theories of gravity could share
similar UV properties.  

Compactification of closed string theories on orbifolds~\cite{orbifolds} -- smooth spaces 
modded out by some discrete group $\Gamma$ -- 
is a classic strategy for constructing four-dimensional models with reduced supersymmetry. The spectrum 
of light (massless) states relevant to the low-energy limit is composed of  {\em untwisted} sector and {\em twisted} 
sector states. The former are the states of ten-dimensional string theory in flat space that are invariant under the action of $
\Gamma$; the latter are the states of strings which would be open in a flat ten-dimensional space but are closed due to the action 
of the orbifold group. Since the mass of a string state is proportional to the length of the string, if the action of $\Gamma$ has 
fixed points (and therefore the orbifold space is singular) the twisted sector has  massless states  localized at the
singularities; otherwise twisted sector states are massive and not directly relevant to the low-energy limit.
However, in all cases twisted sectors are crucial for the consistency of the string construction, in particular for its modular invariance.

Twisted sector states carry charges under the quantum symmetry\footnote{If the orbifold group is Abelian, the quantum 
symmetry group of the orbifold string theory is a group isomorphic to $\Gamma$.} of the orbifold. Thus, in the field theory 
limit, they may be consistently truncated away\footnote{The quantum symmetry guarantees that they appear at least 
bilinearly in the low-energy effective action and thus setting them to zero gives a consistent solution to the nonlinear 
classical equations of motion.}. This leads to a field theory of the $\Gamma$-invariant fields of 
ten-dimensional string theory in flat ten-dimensional Minkowski space, and to an action that is the truncation of the 
effective action of the appropriate ten-dimensional
supergravity\footnote{Alternatively, if a geometric resolution of the (generically asymmetric) orbifold singularities is possible, 
it will generate a nonzero mass for the twisted sector states and thus remove them from the low-energy effective action.}.

The same action may be constructed by starting directly with the ten-dimensional supergravity theory and projecting 
out all states which are not invariant under a discrete Abelian subgroup $\Gamma$ of the $SU(8)$ $R$-symmetry 
group\footnote{This construction guarantees that the truncation of  ten-dimensional supergravity is consistent.}.
Invoking a mild  indulgence of nomenclature, these constructions will be referred to as supergravity orbifolds throughout the paper.
Similar constructions have been discussed extensively for open string theories \cite{orbifold_qfts} where they lead 
to vast classes of quiver gauge theories whose planar limits have special properties.

The UV properties of gravity amplitudes in theories with $\cN \geq 4$ supersymmetry 
have come under recent scrutiny~\cite{Dunbar, Bern:2011rj, Bern:2012cd, Bern:2012gh}. 
In particular, the authors of ref.~\cite{Bern:2012cd} have identified startling\footnote{See refs.~\cite{n4susyAnalysis}  for recent analysis and proposed explanations.}  cancellations 
rendering four-graviton amplitudes in pure $\cN=4 $ supergravity finite through three loops. 
One-loop four-graviton amplitudes in two specific string theory orbifold constructions -- referred to as the 
$(2,2)$ and $(4,0)$ models -- have  been discussed in \cite{Tourkine:2012vx}; very recently it was shown there
that, in the field theory limit,  both the $(2,2)$ and $(4,0)$ orbifold string models lead to the same one-loop 
four-graviton amplitude. It was nevertheless suggested that amplitudes with external vector and scalar fields 
are different in the two constructions. We will find this not to be the case and argue that all one-loop four-point 
amplitudes (and quite likely all-loop all-multiplicity amplitudes) in these theories are the same.

Generically, an orbifold supergravity theory has reduced 
(or no) supersymmetry and has matter supermultiplets in addition to the gravity multiplet.
Matter-coupled supergravity theories, in particular in the presence of vector multiplets, 
are expected to be divergent already at one loop due to 
the existence of a candidate counterterm of the form
\be 
T_M^{\mu \nu} T_{M\mu \nu} \ , 
\ee
where $T_M$ is the matter energy-momentum tensor. 
However, so far explicit computations have been carried out only in theories with ${\cal N}=4$ supersymmetry and 
an arbitrary number of vector multiplets \cite{Fischler:1979yk, Fradkin:1983xs, Bern:2012gh}.

It has been known since the late 1980s that the semi-classical (tree-level) 
S-matrix of Yang-Mills  theory encodes all necessary information to generate the semi-classical S-matrix of gravity.   
This was first demonstrated in the  tree-level relations of Kawai, Lewellen and Tye (KLT)~\cite{KLT} which express gravity 
tree-level amplitudes as the sum over permutations of products  between two color-ordered (or color-stripped) gauge theory amplitudes. 
This double-copy structure was clarified recently by Bern, Johansson, and one of the current authors in~\cite{BCJ}, where 
it was realized that gauge theory amplitudes can be arranged in a graph-organized representation\footnote{Cubic graphs for $D$-dimensional Yang-Mills  theories~\cite{BCJ, BCJLoop}, but quartic graph organizations have made manifest such a duality~\cite{CSm} for (superconformal) Chern-Simons-matter theories in three dimensions.} so as to make manifest a duality between their color and kinematic factors (modulo scalar propagators). Related gravity theories are then trivially given in the same graph organization but with the color factors replaced by another copy of gauge theory kinematic factors.  This duality and its associated double-copy construction are expected to hold at loop level~\cite{BCJLoop} for all gauge theories and related gravity theories that can be organized around scalar graphs. 

The duality between color and kinematics,  as well as the associated double-copy structure of the related gravity theories, such as ${\cal N}=8$ supergravity, 
provide powerful tools for the study of orbifold supergravity 
theories. A particularly 
relevant case is when the orbifold group $\Gamma$ is the product of two groups each acting on a different $SU(4)$ 
factor of the  $SU(4)_R\times SU(4)_L  \subset SU(8)$ symmetry which is manifest in the KLT relations;
such \emph{factorizable} supergravity orbifolds  of  ${\cal N}=8$ supergravity can be expressed as  
double-copies of suitable orbifolds of $\cN=4$ sYM theory.

With this paper, we present a systematic study of one-loop amplitudes in orbifold supergravity theories 
as a concrete step in the search  for supergravity 
theories with reduced supersymmetry that are finite at one loop and beyond.
In particular, we focus on orbifolds of $\cN=8$ supergravity which preserve  $1\leq \cN \leq 4$ supersymmetry 
and have various matter multiplets. We study both gravity and (when applicable) matter amplitudes,  and find 
that all the factorizable orbifold theories with matter have 
divergent one-loop amplitudes when the external  states are taken in the same matter supermultiplet. 
In contrast, pure (matter-free) orbifold supergravity theories with $\cN \leq 4$  are finite at one loop, as expected on general grounds\footnote{As we will see, with the exception of 
$\NeqFour$ supergravity, such  pure reduced supersymmetry theories can be obtained only as non-factorizable 
orbifolds.}. 

The structure of the paper is as follows. In section 2, we review orbifolds of $\cN=8$ 
supergravity and present an exhaustive list of factorizable orbifolds together with some non-factorizable examples.
We work out the maps between  gauge theory and supergravity theory states for the factorizable orbifolds, 
extending some results already in the literature and obtained by other means -- see \cite{Damgaard:2012fb} and 
references therein. 
These maps can be used to construct all loop-level amplitudes using unitarity methods~\cite{UnitarityMethod, BDDK_cut_constructibility} and
the KLT tree-level relations~\cite{KLT},
or directly through the double-copy construction when gauge theory representations that make manifest the duality between 
color and kinematics are available~\cite{BCJ, BCJLoop}. 
For all theories, we identify the moduli space.  We give particular attention to two specific theories, namely $\cN=4$ Maxwell-Einstein
supergravity (MESGT)\footnote{Maxwell-Einstein supergravity theories (MESGT) describe  the coupling of supergravity to vector multiplets.} 
with two vector multiplets and $\cN=2$ Maxwell-Einstein supergravity with seven vector multiplets.
They form a twin pair of supergravity theories that have the same bosonic field content but different number of supersymmetries and hence different fermionic field contents. They can be obtained by consistent truncations of the twin pair  of pure $\cN =6$ supergravity and $\cN=2$ magical MESGT with 15 vector multiplets \cite{GST1}. Both the pure $\cN=6$ supergravity  and the magical MESGT with 15 vector multiplets were obtained as four-dimensional low-energy effective theories of type II superstrings on orbifolds  by 
 Sen and Vafa in their study of dual pairs of compactifications \cite{Sen:1995ff,Gunaydin:2009pk}.  

Magical supergravity theories are a very special class of   $\cN=2$ MESGT whose global symmetry groups in five, four and three dimensions coincide with  the symmetry groups of the famous Magic Square of Freudenthal-Rozenfeld and Tits. In five dimensions they are the only   {\em unified}  MESGTs with symmetric scalar manifolds and are uniquely defined by the four simple Euclidean Jordan algebras  $J_3^{\mathbb{A}}$ of $3\times 3$ Hermitian matrices over the four division algebras $ \mathbb{A}$ (real numbers $ \mathbb{R} $ , complex numbers  $ \mathbb{C} $,
quaternions  $ \mathbb{H} $ and octonions  $ \mathbb{O} $). They describe the coupling of 5 (6), 8 (9), 14 (15) and 26 (27) vector multiplets to $\cN=2$ supergravity 
in five (four) dimensions, respectively~\cite{GST1,GST2}. Of these magical supergravities, the quaternionic theory defined 
by $J_3^{\mathbb{H}}$ has the same bosonic field content as the pure $\cN=6$ supergravity \cite{GST1}. 

Using four-dimensional generalized unitarity \cite{UnitarityMethod, BDDK_cut_constructibility}, in section 3 we construct  
the one-loop four-point amplitudes of the various orbifold theories. In particular, we consider four-graviton amplitudes and 
matter  amplitudes  of the form $F \bar F \rightarrow F \bar F$ and $\phi \bar \phi \rightarrow \phi \bar \phi$.
We find that, after the integral reduction is performed, 
all amplitudes differ from the ones of $\cN=8$ supergravity by additional terms proportional 
to four-dimensional bubble integrals 
and one particular six-dimensional box integral, all multiplied by rational functions in the Mandelstam variables.
For theories that may be obtained by dimensional reduction from six dimensions, we comment on their potential 
additional rational terms from the perspective of the six-dimensional spinor-helicity formalism.

In section 4 we discuss the color/kinematics duality and find representations of one-loop four-point 
amplitudes for gauge theory orbifolds with $\cN=1$ and $\cN=2$ supersymmetry where the duality is manifest.
We demonstrate that the factorized supergravity amplitudes listed in section 2 can be obtained as double-copies of 
suitable gauge amplitudes.  We conclude the paper with a discussion of open questions and directions for further 
research.

%
%
%
%
\renewcommand{\theequation}{2.\arabic{equation}}
\setcounter{equation}{0}

\section{Supergravity orbifolds and truncations of $\cN =8 $ supergravity}

Eleven-dimensional supergravity and type IIB supergravity reduce to ungauged maximal supergravity theories in lower dimensions, $D \leq 9$,  under dimensional reduction, which corresponds simply to compactification  on tori followed by consistent truncation to the massless sector. The resulting maximal supergravity theory in $D$ dimensions has a  global symmetry group $G_D$ and its scalar fields  parametrize the symmetric space 
${M}_D= \frac{G_D}{K_D}$  where $K_D$ is the maximal compact subgroup of $G_D$. Eleven-dimensional 
supergravity reduces to type IIA supergravity in $D=10$ under dimensional reduction. The product manifolds of the $D$-dimensional Minkowski spaces $Min_D$ with the tori $T^{(10-D)}$ provide consistent backgrounds for type II string theories whose low-energy effective theories are simply the maximal supergravity theories in $D$-dimensions. Type IIA and IIB superstring theories compactified over tori are related via T-duality, which in the low-energy limit is only a field redefinition.  The 
eleven-dimensional supergravity theory arises as the low-energy 
effective theory of a strongly-coupled phase of superstring theory~\cite{Witten95}. 

The fields of maximal supergravity theories in various dimensions and their symmetry groups are listed in Table~\ref{maximal}.

Orbifolds, obtained from modding out smooth manifolds by a discrete symmetry group $\Gamma$, can also provide consistent backgrounds for string theories \cite{orbifolds}. Low-energy effective theories of type II superstrings on orbifolds 
of tori $T^{(10-D)}$ can further be consistently truncated -- as field theories -- to the massless fields coming from the {\em untwisted} sector.
The resulting theories are equivalent to consistent  truncations of the $D$-dimensional maximal supergravity to theories 
with fewer supersymmetries and fields that are invariant under the action of $\Gamma$. We will refer to them as {\em supergravity orbifolds}, in analogy with the similar gauge theory constructions \cite{orbifold_qfts}.
%

Our main goal is to analyze -- from the perspective of their scattering amplitudes -- a large set of supergravity theories 
that are orbifold projections of  $\NeqEight$ supergravity. We will use the KLT relations to describe the tree-level 
amplitudes of these theories as sums of products of color-stripped tree-level 
amplitudes of sYM theories with less-than-maximal supersymmetry.
The map of states relevant to maximal supergravity \cite{KLT} uses the $SU(4)_L\times SU(4)_R$ basis of the R-symmetry
group $SU(8)$ of four-dimensional maximal supergravity. Each $SU(4)$ factor is  identified with the R-symmetry
group of an $\NeqFour$ sYM theory. 
The orbifold projection may be chosen to act either independently on the two factors or in a 
correlated fashion. As usual, we will refer to them as {\em factorizable} and {\em non-factorizable} orbifolds, respectively,  
and focus mainly on the factorizable ones. 
%
There exists a limited set of relevant gauge theories which can be obtained as orbifold projections of ${\cal N}=4$ sYM theory 
by some discrete group $\Gamma$ and have only fields in the adjoint representation of $SU(N)$\footnote{Of course, there exists a wide choice of orbifold theories which have fields in bifundamental 
representation \cite{orbifold_qfts}; the main difference between those orbifolds and the ones discussed here is that {\em here} the action of 
the orbifold group on the gauge degrees of freedom is taken to be trivial.}:
\begin{enumerate}
\item A unique $\Gamma={\bf Z}_2$ orbifold preserving ${\cal N}=2 $ supersymmetry where $\Gamma$ is generated by  
 diag$(1,1,-1,-1)$. The theory contains one vector, two fermions and two real scalars or, equivalently, one ${\cal N}=2$ vector multiplet.

\item A unique $\Gamma={\bf Z}_3$ orbifold preserving ${\cal N}=1$ supersymmetry where $\Gamma$ is generated by
diag$(1, e^{{2\over 3} \pi  i }, e^{{2\over 3} \pi  i } ,e^{{2\over 3} \pi  i })$. The theory contains one vector, 
one fermion and no scalars, that is a single ${\cal N}=1$ vector multiplet.

\item A non-supersymmetric $\Gamma={\bf Z}_2$ orbifold generated by $-I_4$. The theory contains one vector, six real scalars and no fermions.

\item A non-supersymmetric $\Gamma={\bf Z}_2 \times {\bf Z}_2$ orbifold
generated by  diag$(1,1,-1,-1)$ and diag$(-1,-1,1,1)$. This theory has one vector, two real scalars and no fermions.

\item A non-supersymmetric $\Gamma={\bf Z}_4$ orbifold generated by $e^{{ \pi\over 2} i}I_4$. This theory contains one vector and no other fields.

\item A non-supersymmetric $\Gamma={\bf Z}_4$, which we shall denote as ${\bf \tilde{Z}}_4$ orbifold, generated by ${\rm diag}(i, -i ,-i, i)$. 
This theory contains one vector, four real scalars and no fermions.

\end{enumerate}

{
\renewcommand{\arraystretch}{1.1}
\begin{table}[t]
\small
\begin{center}
\begin{tabular}{|c|c|c|}
\hline
{\bf Dimension }  & {\bf Field Content} & ${M}_D = G_D / K_D $ \\[5pt] 
\hline
$D=6 $ &   $g_{\mu\nu}, 4 \psi_{\mu}, 5 B_{\mu\nu}, 16 A_\mu, 20  \lambda_{}, 25 \phi $ & ${SO(5,5) / \big(SO(5)\times SO(5)}\big)$\vphantom{${}^\big{|}$} \\[5pt]
\hline
$D=5 $ &   $g_{\mu\nu}, 8\psi_\mu, 27 A_\mu, 48 \lambda, 42 \phi $ & $E_{6(6)} /USp(8)$\vphantom{${}^\big{|}$} \\[5pt]
\hline
$D=4$ & $g_{\mu\nu}, 8\psi_\mu, 28 A_\mu, 56 \lambda, 70 \phi $ & $E_{7(7)} /SU(8) $\vphantom{${}^\big{|}$} \\[5pt]
\hline
$D=3 $ & $128 \lambda, 128 \phi $ & $ E_{8(8)}/SO(16)$\vphantom{${}^\big{|}$} \\[5pt]
\hline
\end{tabular}
\caption{ \small Fields of maximal Poincar\'e supergravity in various dimensions and their scalar manifolds\label{maximal}.}
\end{center}
\end{table}}

In Table~\ref{taborb} we list  factorizable orbifold groups of the form ${\bf Z_L} \times {\bf Z_R}$, where 
${\bf Z_L} ({\bf Z_R})$ is a discrete subgroup of $SU(4)_L (SU(4)_R)$, together with the massless field content  of the corresponding supergravity  orbifolds. 
\footnote{By taking non-supersymmetric orbifold groups for both ${\bf Z_L}$ and ${\bf Z_R}$ 
one could also construct non-supersymmetric gravity theories with various amounts of matter. We will however restrict 
ourselves to supergravity theories with at least $\NeqOne$ supersymmetry.}

Other choices of ${\bf Z_L}$ and ${\bf Z_R}$ do not lead to additional new theories as long as their action on the gauge indices of  the fields of $\cN=4$ sYM   
is trivial; therefore, the orbifold supergravities in Table~\ref{taborb} are all the factorizable ones that preserve at least minimal amount of supersymmetry.

As an example, we work out the massless field content of the $\cN =4$ theory with two vector multiplets on the fifth row of Table \ref{taborb}. 
For this theory, the  ${\bf Z}_2 \times {\bf Z}_2$ orbifold 
group is generated by the elements\footnote{Let us note that, from the perspective of the $SU(8)$ R-symmetry group of
$\NeqEight$ supergravity, the position of the negative entries in the generator of ${\bf Z}_2\times {\bf Z}_2$ is given by 
a particular choice of labels of the $SU(8)$ Cartan generators, which is completely arbitrary.}
 diag$(1,1,-1,-1,1,1,1,1)$ and diag$(1,1,1,1,1,1,-1,-1)$. Aside from the graviton, which is unaffected by the projection, 
four gravitini, namely $\psi^{a}_\mu$ with $a=1,2,5,6$, are left in the orbifold theory. 
The theory also has six vector fields which are part of the gravity multiplet, $A_{\mu}^{ab}$ with $a,b=1,2,5,6$, and two additional matter vectors, $A^{34}_{\mu}$ and 
$A^{78}_{\mu}$. Moreover, a total of twelve spin-${1 \over 2}$ fields survive the projection, namely $\l^{abc}$, $\l^{a34}$ and $\lambda^{a78}$ with $a,b,c=1,2,5,6$. 
The theory is completed by a total of fourteen real scalars of the form $\phi^{1256},\phi^{ab34},\phi^{ab78}$ and $\phi^{3478}$ with $a,b=1,2,5,6$. 
It is immediate to verify that all other fields are not invariant under at least one of the two generators of the orbifold group.
{ 
\renewcommand{\arraystretch}{1.25}

\begin{table}[H]
\small
\begin{center}
\begin{tabular}{|c|c|c|c|c|}
\hline
{\bf $\cal N$ }&{\bf Orbifold } & {\bf Factors}  & {\bf Massless Fields} & {\bf  Matter}\\ 
\hline

$6$ & $ {\bf Z}_2$  & $ \cN=4 \times \cN=2|_{{\bf Z}_2}$ & $g_{\mu\nu}, 6\psi_\mu, 16 A_\mu, 26 \lambda, 30 \phi $ & pure \\
\hline
$5$ & ${\bf Z}_3$& $ \cN=4 \times \cN=1|_{{\bf Z}_3}$ & $g_{\mu\nu}, 5\psi_\mu, 10 A_\mu, 11 \lambda, 10 \phi $ & pure \\
\hline
$4$ & ${\bf Z}_2$ & $ \cN=4 \times \cN=0|_{{\bf Z}_2}$ &  $g_{\mu\nu}, 4\psi_\mu, 12 A_\mu, 28 \lambda, 38 \phi $ & $6$ vectors \\
$4$ & $ {\bf \tilde{Z}}_4$ & $ \cN=4 \times \cN=0|_{\tilde {\bf Z}_4}$  & $g_{\mu\nu}, 4\psi_\mu, 10 A_\mu, 20 \lambda, 26 \phi $ & $4$ vectors  \\
$4$ & ${\bf Z}_2\times{\bf Z}_2$ & $ \cN=4 \times \cN=0|_{{\bf Z}_2\times {\bf Z}_2}$ &  $g_{\mu\nu}, 4\psi_\mu, 8 A_\mu, 12 \lambda, 14 \phi $ & $2$ vectors \\
$4$ & ${\bf Z}_2\times{\bf Z}_2$ & $ \cN=2|_{{\bf Z}_2} \times \cN=2|_{{\bf Z}_2}$ &  
$g_{\mu\nu}, 4\psi_\mu, 8 A_\mu, 12 \lambda, 14 \phi $ & $2$ vectors \\
$4$ & ${\bf Z}_4$& $\cN=4\times\cN=0|_{{\bf Z}_4}$ & $g_{\mu\nu}, 4\psi_\mu, 6 A_\mu, 4 \lambda, 2 \phi $ & pure \\
\hline
$3$ & ${\bf Z}_2\times {\bf Z}_3$ & $ \cN=2|_{{\bf Z}_2} \times \cN=1|_{{\bf Z}_3}$ &$g_{\mu\nu}, 3\psi_\mu, 4 A_\mu, 5 \lambda, 6 \phi $ &
$1$ vector \\
\hline
$2$ & ${\bf Z}_2\times {\bf Z}_2$ & $ \cN=2|_{{\bf Z}_2} \times \cN=0|_{{\bf Z}_2}$ &$g_{\mu\nu}, 2\psi_\mu, 8 A_\mu, 14 \lambda, 14 \phi $ & $7$ vectors \\

$2$ & ${\bf Z}_2 \times {\bf \tilde{Z}}_4$ & $ \cN=2|_{{\bf Z}_2} \times \cN=0|_{\tilde {\bf Z}_4}$ &  $ g_{\mu\nu}, 2\psi_\mu, 6 A_\mu, 10 \lambda, 10 \phi $ & $5$ vectors \\

$2$ & ${\bf Z}_2\times {\bf Z}_2\times {\bf Z}_2$ & $ \cN=2|_{{\bf Z}_2} \times \cN=0|_{{\bf Z}_2\times{\bf Z}_2}$ &
$g_{\mu\nu}, 2\psi_\mu, 4 A_\mu, 6 \lambda, 6 \phi $ & $3$ vectors \\
$2$ & ${\bf Z}_2\times {\bf Z}_4$ & $ \cN=2|_{{\bf Z}_2} \times \cN=0|_{{\bf Z}_4}$ & $g_{\mu\nu}, 2\psi_\mu, 2 A_\mu, 2 \lambda, 2 \phi $ & 
$1$ vector \\
$2$ & ${\bf Z}_3\times {\bf Z}_3$ & $ \cN=1|_{{\bf Z}_3} \times \cN=1|_{{\bf Z}_3}$ & $g_{\mu\nu}, 2\psi_\mu, 1 A_\mu, 2 \lambda, 4 \phi $ & 
$1$ hyper \\
\hline

$1$ & ${\bf Z}_3\times {\bf Z}_2$ & $ \cN=1|_{{\bf Z}_3} \times \cN=0|_{{\bf Z}_2}$ &  $g_{\mu\nu}, 1\psi_\mu, 6 A_\mu, 7 \lambda, 2 \phi $ &
$6$ vecs, $1$ chiral \\

$1$ & ${\bf Z}_3 \times {\bf \tilde{Z}}_4$ & $ \cN=1|_{{\bf Z}_3} \times \cN=0|_{\tilde {\bf Z}_4}$ & $ g_{\mu\nu}, \psi_\mu, 4 A_\mu, 5 \lambda, 2 \phi $ & $4$ vecs, $1$ chiral  \\

$1$ & ${\bf Z}_3\times {\bf Z}_2\times  {\bf Z}_2$ & $ \cN=1|_{{\bf Z}_3} \times \cN=0|_{{\bf Z}_2\times{\bf Z}_2}$ &
$g_{\mu\nu}, 1\psi_\mu, 2 A_\mu, 3 \lambda, 2 \phi $ & 
$2$ vecs, $1$ chiral \\
$1$ & ${\bf Z}_3\times {\bf Z}_4$ & $ \cN=1|_{{\bf Z}_3} \times \cN=0|_{{\bf Z}_4}$ & $g_{\mu\nu}, 1\psi_\mu,  1 \lambda, 2 \phi $ & 
$1$ chiral \\

\hline
\end{tabular}
\caption{ \small Factorizable orbifold groups and  massless field content of the low-energy four-dimensional 
supergravities coming from the untwisted sectors of the corresponding string theory orbifolds.
The list includes all the supergravity theories that can be obtained though a factorized orbifold group action. 
The scalar fields in the fourth column are real.\label{taborb}}
\end{center}
\end{table}}

In Table~\ref{taborb_moduli}  we list the number of supersymmetries coming from left and right  
sectors of four-dimensional orbifold supergravity  theories and the corresponding moduli spaces, which  are all submanifolds of $E_{7(7)}/SU(8)$. 

We should  note that there also exist  orbifold theories which are non-factorizable. 
These theories are largely outside of the scope of this paper. A partial list can be found in Table~\ref{taborb2}. 


{ \renewcommand{\arraystretch}{1.25}
\begin{table}[H]
\small
\begin{center}
 
 \vspace{-5pt}
 
\begin{tabular}{|c|c|c|c|c|}
\hline
\# & {\bf $\cal N$}& {\bf Factor}s  & {\bf Scalar Manifold} & {\bf Supergravity} \\ 
\hline

1 & $6$ &  ${\cal N}=4  \times  {\cal N}=2|_{{\bf Z}_2}  $ & $ \frac{SO^*(12)}{U(6)}$
\vphantom{${}^{|^{{}^A}}$}  & {\small pure ${\cal N}=6$ sugra} \\[2pt]
\hline
2 & $5$ &  $ {\cal N}=4  \times  {\cal N}=1|_{{\bf Z}_3}   $ &$ \frac{SU(5,1)}{U(5)}$ 
\vphantom{${}^{|^{{}^A}}$}& {\small pure ${\cal N}=5$ sugra} \\[2pt]
\hline
3 &  $4$ &  ${\cal N}=4 \times {\cal N}=0|_{{\bf Z}_2}  $ & $ \frac{SO(6,6)\times SU(1,1)}{SO(6)\times SO(6) \times U(1)}$ 
& \small{\tabentry{${\cal N}=4$ sugra,}{6 vector multiplets}} \\[0pt]
4& $4$ &  ${\cal N}=4 \times {\cal N}=0|_{{\bf \tilde{Z}}_4}  $ & $ \frac{SO(6,4)\times SU(1,1)}{ U(4)\times SO(4) }$ & 
\small{\tabentry{${\cal N}=4$ sugra,}{4 vector multiplets}} \\[0pt]
5 & $4$ &  ${\cal N}=4 \times {\cal N}=0|_{{\bf Z}_2\times {\bf Z}_2} $ & $ \frac{SO(6,2)\times SU(1,1)}{SO(6)\times SO(2) \times U(1)}$  
& \small{\tabentry{${\cal N}=4$ sugra,}{$2$ vector multiplets}} \\[0pt]
6& $4$ &  ${\cal N}=2|_{{\bf Z}_2} \times {\cal N}=2|_{{\bf Z}_2}  $ & 
$ \frac{SO(6,2)\times SU(1,1)}{SO(6)\times SO(2) \times U(1)}$   
& \small{\tabentry{${\cal N}=4$ sugra,}{$2$ vector multiplets}} \\[0pt]
7& $4$ &  ${\cal N}=4 \times {\cal N}=0|_{{\bf Z}_4}  $ & $ \frac{SU(1,1)}{ U(1)}$ & {\small pure ${\cal N}=4$ sugra} \\
&&&& \\[-14pt]
\hline
8& $3$ &  ${\cal N}=2|_{{\bf Z}_2} \times {\cal N}=1|_{{\bf Z}_3}  $ & $ \frac{SU(3,1)}{U(3)}$  
& \small{\tabentry{${\cal N}=3$ sugra,}{$1$ vector multiplet}} \\[0pt]
\hline
9 & $2$ &  ${\cal N}=2|_{{\bf Z}_2} \times {\cal N}=0|_{{\bf Z}_2}  $ & 
$ \frac{SO(6,2)\times SU(1,1)}{SO(6)\times SO(2) \times U(1)}$   
& \small{\tabentry{${\cal N}=2$ sugra,}{$7$ vector multiplets}} \\[0pt]
10 & $2$ &  ${\cal N}=2|_{{\bf Z}_2} \times {\cal N}=0|_{{\bf \tilde{Z}}_4}  $ & 
$ \frac{SO(4,2)\times SU(1,1)}{SO(4)\times SO(2) \times U(1)}$    
& \small{\tabentry{${\cal N}=2$ sugra,}{$5$ vector multiplets}} \\[0pt]
11 & $2$ &  ${\cal N}=2|_{{\bf Z}_2} \times {\cal N}=0|_{{\bf Z}_2\times{\bf Z}_2} $ &
$ \frac{SU(1,1)\times SU(1,1)\times SU(1,1)}{U(1)\times U(1)  \times U(1)}$  
& \small{\tabentry{${\cal N}=2$ sugra,}{$3$ vector multiplets}} \\[0pt]
12 & $2$ &  ${\cal N}=2|_{{\bf Z}_2} \times {\cal N}=0|_{{\bf Z}_4}  $ & $ \frac{ SU(1,1)}{ U(1)}$  
& \small{\tabentry{${\cal N}=2$ sugra,}{$1$ vector multiplet}} \\[0pt]
13 & $2$ &  ${\cal N}=1|_{{\bf Z}_3} \times {\cal N}=1|_{{\bf Z}_3}  $ & $ \frac{SU(2,1)}{SU(2) \times U(1)}$  
& \small{\tabentry{${\cal N}=2$ sugra,}{$1$ hypermultiplet}} \\[0pt]
\hline
14 & $1$ &  ${\cal N}=1|_{{\bf Z}_3} \times {\cal N}=0|_{{\bf Z}_2}  $ & $ \frac{SU(1,1)}{ U(1)}$   
& \small{\tabentry{$\cN =1$ sugra, $6$ vector}{ and $1$ chiral multiplets}} \\[0pt]
15 & $1$ &  ${\cal N}=1|_{{\bf Z}_3} \times {\cal N}=0|_{{\bf \tilde{Z}}_4}  $ & $ \frac{SU(1,1)}{ U(1)}$ 
& \small{\tabentry{$\cN =1$ sugra, $4$ vector}{ and $1$ chiral multiplets}} \\[0pt]
16 & $1$ &  ${\cal N}=1|_{{\bf Z}_3} \times {\cal N}=0|_{{\bf Z}_2\times{\bf Z}_2}  $  & $ \frac{SU(1,1)}{ U(1)}$ 
& \small{\tabentry{$\cN =1$ sugra, $2$ vector}{ and $1$ chiral multiplets}} \\[0pt]
17 & $1$ &  ${\cal N}=1|_{{\bf Z}_3} \times {\cal N}=0|_{{\bf Z}_4}  $ & $ \frac{SU(1,1)}{ U(1)}$ 
& \small{\tabentry{$\cN =1$ sugra,}{$1$ chiral multiplets}} \\[0pt]
\hline
\end{tabular}

\end{center}

\caption{ \small Factorizable ${\cal N}\ge 1$  supergravity orbifolds and their corresponding scalar manifolds. The
second column lists the total number of preserved $D=4$ supersymmetry.  The third column lists the number of supersymmetries  coming from the different gauge factors. 
The fifth column lists the supergravity theories and the fourth column their scalar manifolds. 
Note that the theories with ${\cal N}=2,3,4$ are ungauged Maxwell-Einstein Supergravity Theories (MESGTs).
\label{taborb_moduli}}
\end{table}}


 { 
 \renewcommand{\arraystretch}{1.1}
 \begin{table}[t]
 \small
 \begin{center}
\begin{tabular}{|c|cc@{}cc|}
\hline
{\bf $\cN$} &{\bf Orbifold} & {\bf Generators} & {\bf Field Content} &{\bf Matter }\\ 
\hline

$3$ & ${\bf Z}_5$ &$I_3 \times e^{{2\over 5} \pi i }I_5$ & $g_{\mu\nu}, 3\psi_\mu, 3 A_\mu, 1 \lambda $ &
pure \\
\hline
$2$ & ${\bf Z}_6$ & $I_2\times e^{{1\over 3}  \pi i} I_6$ & $g_{\mu\nu}, 2\psi_\mu, 1 A_\mu $ &
pure \\
$2$ & ${\bf Z}_3$ & $I_2\times e^{{2\over 3}  \pi i} I_6$ & $g_{\mu\nu}, 2\psi_\mu, 1 A_\mu, 20 \lambda, 40 \phi $ &
$10$ hypers \\
$2$ & ${\bf Z}_2$ & $I_2\times e^{  \pi i} I_6$& $g_{\mu\nu}, 2\psi_\mu, 16 A_\mu, 30 \lambda, 30 \phi $ &
$15$ vectors \\
\hline
$1$ & ${\bf Z}_7$ & $I_1\times e^{{2\over 7}  \pi i} I_7$ & $g_{\mu\nu}, 1\psi_\mu $ &
pure \\
$1$ & ${\bf Z}_5 \times {\bf Z}_2$ & $I_1\times ( - I_2) \times I_5; \ I_3 \times e^{{2 \over 5}  \pi i} I_5  $
& $g_{\mu\nu}, 1\psi_\mu, 1 A_\mu, 1 \lambda  $ &
$1$ vector \\
\hline
\end{tabular}
\caption{ \small Examples of non-factorizable supergravity orbifolds preserving some supersymmetry. The third column lists 
the orbifold group generators and the $SU(8)$ indices they act on, up to conjugation; 
for example, $I_3 \times e^{{2\over 5} \pi i }I_5$
generates a ${\bf Z}_5$ group that acts on five of the eight values of an $SU(8)$ fundamental representation index. 
The ${\cal N}=2$ theory  with $10$ hypermultiplets has the moduli space $\frac{E_{6(2)}}{SU(6)\times SU(2)}$
and the $\cN=2$ theory  with $15$ vector multiplets has the same moduli space as the $\cN=6$ supergravity, namely 
$\frac{SO^*(12)}{U(6)}$.  \label{taborb2}}
 \end{center}
\end{table}
}

Dual pairs of type II string compactifications on orbifolds were studied by Sen and Vafa \cite{Sen:1995ff}, who 
constructed examples of dual pairs with $\cN=2$, $\cN= 4$ and $\cN=6$ supersymmetry. Among the 
four-dimensional $\NeqTwo$ theories obtained in this way, there is one 
with fifteen vector multiplets. As it was pointed out in \cite{Gunaydin:2009pk}, this $\NeqTwo$ MESGT is 
simply  the $\NeqTwo$ magical supergravity theory defined by the
Jordan algebra $J_3^{\mathbb{H}}$ of $3\times 3$ Hermitian matrices over the quaternions \cite{GST1,GST2}. 
This quaternionic $\NeqTwo$ magical supergravity
and $\cN=6$ supergravity 
have the same bosonic field content with  the   scalar manifold $SO^*(12)/U(6)$, and hence form a twin pair of 
supergravities~\cite{GST1}\footnote{ Both the quaternionic magical MESGT and the $\cN=6$ supergravity 
are self-dual as type II compactifications in the
sense of S and T duality exchange \cite{Sen:1995ff}.}. $\cN=6$ supergravity is obtained as a factorized orbifold  
(the first row  in Table~\ref{taborb_moduli}) while
the quaternionic magical supergravity is obtained as a non-factorizable orbifold (fourth row in Table~\ref{taborb2}). 
We should also note that under dimensional reduction  $\cN=6$ supergravity and 
the magical quaternionic MESGT yield twin supergravities  in three dimensions with $\cN=12$  
and $\NeqFour$ supersymmetries  and  the same scalar manifold, namely $\frac{E
 _{7(-5)}}{SO(
12)\times SU(2)}$. 

The $\cN=6$ supergravity and the magical supergravity defined by $J_3^{\mathbb{H}}$ can be truncated to $\NeqFour$ and $\NeqTwo$
MESGTs  that also form a twin pair with the  same bosonic field content. The resulting twin theories 
are $\NeqFour$ supergravity coupled to two 
vector multiplets and $\NeqTwo$ supergravity coupled to seven 
vector multiplets and have the same scalar manifold, $\frac{SO(6,2)\times SU(1,1)}{U(4)\times U(1)}$. The  $\NeqFour$ MESGT with two
vector multiplets can be obtained as
a factorized orbifold theory in two different ways,  which are listed in rows five and six of  Table~\ref{taborb_moduli}. 
 The corresponding twin  $\NeqTwo$ MESGT with seven vector multiplets is listed in row nine of the same table.
When reduced to three dimensions they yield a twin pair of  $\NeqEight$ and $\NeqFour$ supergravity theories, 
respectively,  with the same scalar manifold $\frac{SO(8,4)}{SO(8)\times SO(4)}$. 

We should note that the magical $\NeqTwo$ supergravity defined by the quaternionic Jordan algebra $J_3^{\mathbb{H}}$ can be further 
truncated to the magical supergravity theory defined by complex Jordan algebra $J_3^{\mathbb{C}}$ with nine vector
multiplets and scalar manifold $\frac{SU(3,3)}{SU(3)\times SU(3)\times U(1)}$  and to the magical supergravity defined by real  
Jordan algebra $J_3^{\mathbb{R}}$ with six vector multiplets and scalar manifold $\frac{Sp(6,\mathbb{R})}{U(3)}$.
None of the magical supergravity theories involves any hypermultiplets. 

One can go further and truncate the smallest magical supergravity theory to the STU model which describes the coupling 
of three  $\NeqTwo$ vector multiplets to supergravity in four dimensions. The STU model has the global symmetry 
group $SU(1,1)^3$  under which the vector field strengths and their magnetic duals transform in the 
eight-dimensional representation $(1/2,1/2,1/2)$. 
The STU model coupled to  four hypermultiplets is one of the theories obtained by Sen and Vafa by orbifolding \cite{Sen:1995ff}. 

In light of the above comments, we should stress that every toroidal orbifold compactification of type II superstring theory 
leads to a supergravity orbifold of maximal supergravity involving the massless fields of the untwisted sector. 
At the supergravity level one can consistently truncate these theories further by discarding certain fields. 
However, the truncated theory may generally not be the low-energy effective theory of the massless untwisted sector of some  orbifold of type II superstring. %
For example, ref.~\cite{Dabholkar:1998kv} lists  pure supergravity theories in various dimensions that cannot be obtained by toroidal orbifolds from M-theory. 
%
%
They can however all be obtained by consistent truncation of the maximal supergravity in the respective dimension. 
Among the supergravity orbifolds we have listed in Table~\ref{taborb_moduli} there is a  $\NeqTwo$ MESGT with only  one $\NeqTwo$ vector multiplet. 
This theory can be obtained by dimensional reduction from the pure $\NeqTwo$ supergravity in five dimensions which has no moduli.
Elaborating further on the results of  \cite{Dabholkar:1998kv}, it was argued in \cite{Mizoguchi:2001cp} that the moduli-free pure $\NeqTwo$ 
supergravity in five dimensions cannot be obtained from type II superstring by orbifolding and that any such orbifolding  
must involve additional multiplets containing some moduli. This implies that there must be additional massless fields 
coming from the twisted sector of the corresponding  orbifold.


%
%
%
%
\renewcommand{\theequation}{3.\arabic{equation}}
\setcounter{equation}{0}

\section{One-loop amplitudes at four points \label{direct_calculation}}

We now turn to the calculation of one-loop amplitudes in the factorizable orbifold theories listed in Table~\ref{taborb_moduli} and in the non-factorizable ones listed in Tables~\ref{taborb2}; we will use the
generalized unitarity method, originally introduced in \cite{UnitarityMethod}, \cite{BDDK_cut_constructibility} and further
developed for one-loop amplitudes in \cite{BCFUnitarity}. 
We will evaluate unitarity cuts in four dimensions and thus our results could potentially miss\footnote{This is to be contrasted with constructing supersymmetric gauge theory amplitudes at one loop, where improved power 
counting guarantees~\cite{BDDK_cut_constructibility} 
 that four-dimensional unitarity methods
yield a complete result through ${\cal O}(\epsilon^0)$ in the integrated expression.} rational functions of momentum invariants. 
In cases where  $D$-dimensional unitarity cuts are readily accessible
(e.g. ${\cal N}=4$ supergravity coupled with two vector multiplets) we will nevertheless see that our results are complete 
through ${\cal O}(\epsilon^0)$.

\subsection{On tree-level amplitudes and unitarity cuts}

In analogy to gauge theory orbifolds \cite{orbifold_qfts}, the fields of orbifold supergravity theories can be obtained by suitably projecting the fields of the $\NeqEight$ supergravity parent theory.
Assuming that the orbifold group is Abelian and denoting it  by $\Gamma$,  the relevant projection operators are 
\be 
{\cal P}_\Gamma \ \Phi_{a_1 \ldots a_n } = {1 \over | \Gamma  |} \sum_{r \in \Gamma} 
r^{a_1}_{a_1} \ldots r^{a_n}_{a_n}  \Phi_{a_1 \ldots a_n} \ , \label{proj1} 
\ee
where $\Phi_{a_1 \ldots a_n}$ is a generic field of ${\cal N}=8$ supergravity labeled by the fundamental $SU(8)$ indices 
$a_1 \ldots a_n = 1,2,\ldots,8$ and the diagonal matrices $r$ form a representation of $\Gamma$ and act on 
the fundamental representation of $SU(8)$. This projection operator may also be interpreted as acting on the 
Grassmann parameters labeling the ${\cal N}=8$ on-shell superfield; it may therefore be used to extract from the 
${\cal N}=8$ tree-level superamplitudes the component amplitudes of the orbifolded theory. An $n$-point superamplitude
in the orbifold supergravity theory is thus given  as 
\be 
{\cal M}^\text{tree}_{n,\Gamma} = {1 \over |\Gamma|} \sum_{r_{1} \in \Gamma} \ldots {1 \over |\Gamma|} \sum_{r_{n} 
\in \Gamma}  {\cal M}^\text{tree}_n \Big((r_{1})^a_a \eta_{(1)}^a, \ldots, (r_{n})^a_a \eta_{(n)}^a \Big) \ . 
\label{treesuperamp}
\ee
This identification of tree superamplitudes implies that we can carry out the sum over states that cross (generalized) 
unitarity cuts (or supersums) as a Grassmann  integral using familiar techniques  developed  for ${\cal N}=4$ sYM 
 and ${\cal N}=8$  supergravity \cite{Elvang:2008na, Bern:2009xq, ArkaniHamed:2008gz, Bianchi:2008pu, Lal:2009gn, Elvang:2011fx}. 
For example, the $s$-channel two-particle cut of  the four-point one-loop superamplitude is 
\be 
\left. {\cal M}_{4,\Gamma}^{(1)}(1,2,3,4)\right|_{s-\mathrm{cut}} \!\!\!= \!\Big({\kappa \over 2}\Big)^4 \sum_{r \in \Gamma} { s^2 \ 
\delta^8 \Big(\sum^2_{k=1} \langle k l_1 \rangle \eta^a_k + r^a_a \sum^4_{k=3} \langle k l_1 \rangle \eta^a_k \Big)
\delta^8 \Big( \sum_{k=1}^4 \langle k l_2 \rangle \eta^a_k \Big)
\over  \langle 12 \rangle^2 \langle 34 \rangle^2 \langle l_1 l_2 \rangle^4 
\prod^4_{k=1} \langle k l_1 \rangle \langle k l_2 \rangle}  \ . \label{oneloopsuperamp} 
\ee
%
The $s$-channel cuts of the component amplitudes can be obtained by acting with an appropriate 
set of Grassmann derivatives on  the expression (\ref{oneloopsuperamp}). 
Similarly, the $t$- and $u$-channel cuts at one loop can be obtained by relabeling the external legs.

We proceed to construct the one-loop amplitudes as follows:
\begin{enumerate}
 
 \item We start with an \Ansatz~ involving box and triangle 
 scalar integrals multiplied by
 rational functions of the Mandelstam variables and of the scalar products $\tau_i=k_i \cdot l$, where $l$ is the loop momentum. 
 The degrees of the polynomials appearing in the \Ansatz~  depend on the number of supersymmetries 
 of the particular orbifold theory we are considering. 

 \item We fix most of the coefficients in the \Ansatz~  by requiring that its $s$-, $t$- and $u$-channel two-particle cuts 
 match the direct evaluation of these cuts in terms of tree-level (super)amplitudes (cf. eq.~(\ref{oneloopsuperamp})).
 At this stage, some of the coefficients in the \Ansatz~  are still free; some of them correspond to tadpoles and 
 bubble-on-external-line (snail) integrals\footnote{Such contributions should be absent from supersymmetric scattering
 amplitudes. They appear however because we made no effort to remove then at the previous step.}.
 
 \item We perform a Passarino-Veltman reduction using the expressions collected in appendix~\ref{appPV}. 
 Due to the presence of supersymmetry, tadpole and bubble-on-external-line integrals should not be present 
 at this loop level;
 while they may emerge from the integral reduction due to the features of the original \Ansatz, they are discarded.

\end{enumerate}

After the reduction is carried out,  all the remaining free coefficients disappear and the results 
assume remarkably simple expressions,  which are listed in the following subsections. Our results are given up 
to ${\cal O}(\epsilon) $ terms and up to rational terms. Using the six-dimensional spinor helicity formalism 
\cite{Dennen:2009vk} we comment in  sec.~\ref{subsecrational}  on the properties of the rational terms in some of these 
theories (which, in cases that we have checked, turn out to be complete).

\subsection{Four-graviton amplitudes \label{subsecgrav}}

In this subsection we focus on four-graviton amplitudes ${\cal M}(--++)$ for the 
various orbifold theories listed in Table~\ref{taborb}. 
These amplitudes are expected to be finite regardless of the theory, since the first candidate on-shell counterterm  
involving four gravitons arises only at three loops.
The amplitudes are constructed using an \Ansatz~  of the form\footnote{Parity-odd terms, which integrate to zero, are ignored.},
\be 
{\cal M}^{(1)}_{4} =\Big({\kappa \over 2}\Big)^4 {{M}_{\textrm{tree}} \over s^{7-{\cal N}}} { {\int {d^{4-2 \epsilon} l \over (2 \pi)^{4-2 \epsilon}}}} 
\Big( \sum_{n=1}^3 P^{(n)}_{{10-{\cal N}},0} 
\big(s_{ij}\big) {  \cal I}_{4,n}^{D=4-2\epsilon}+ \sum_{m=1}^6 Q^{(m)}_{{9-{\cal N}},N_L} 
\big(s_{ij}, \tau_k \big) {  \cal I}^{D=4-2\epsilon}_{3,m}  \Big) \ , \label{ansatzN4}
\ee 
where ${\cal N}$ is the number of supersymmetries and  
\be 
{ M}_{\mathrm{tree}} =i {s t \over u } A_{\mathrm{tree}}^2 = - i s {\langle12 \rangle^6 \over \langle 34 \rangle^2 \langle13 \rangle \langle14 \rangle \langle23 \rangle
\langle24 \rangle } \ , 
\ee
with $A_{\mathrm{tree}}$ being the tree-level four-gluon color-ordered amplitude.
$P^{(n)}_{p,q}$ and $Q^{(m)}_{p,q}$ are generic polynomials of degree 
$p$ in the Mandelstam variables 
$\{s_{12},s_{23},s_{13}\}=\{ s,t,u\}$, with $s_{ij}\equiv (k_i+k_j)^2$, 
and in the scalar products $\tau_i = k_i \cdot l $ with  terms up to degree $q$ in the loop momentum~$l$. 
${  \cal I}_{4,n}^{D=4-2\epsilon}$ and
${  \cal I}_{3,m}^{D=4-2\epsilon}$, with $n=1,2,3$ and $m=1,\ldots,6$, are 
the three independent scalar box {integrands} and the six independent (before integration) scalar triangle {integrands}, 
respectively. 
The four-point amplitudes in these theories do not require that explicit bubble integrands be present in the \Ansatz.  Nevertheless,  
bubble integrals can (and in some cases do) appear after the reduction to an integral basis is carried out.   
All bubbles-on-external-legs are discarded. 
The maximum number of loop momenta in the numerators of the triangle integrals, $N_L$, is chosen to be
\be 
N_L = 6 - \cN, \qquad  \cN= 2, 3, 4 \ .  
\ee
Interestingly, this pattern is improved for $\cN = 1$, in which case it is sufficient to have four loop momenta in the numerators.

\begin{table}[t]
\small
\begin{center}
 
{ \renewcommand{\arraystretch}{1.4}

\begin{tabular}{|c|ccccc|}
\hline
{\bf$\cN$}& {\bf Orbifold}  & {\bf Matter} & $ {\cal P}_1 (\a)$  
& $ {\cal P}_2 (\a)$ & $ {\cal Q}(\a)$   \\[2pt] 
\hline

$4$ & ${\cal N}=4 \times {\cal N}=0|_{{\bf Z}_2}  $  & 6 vecs & $ 4 $ & $ 4 - 8 {\a}$ & $4$ \\

$4$ & ${\cal N}=4 \times {\cal N}=0|_{\tilde {\bf Z}_4}  $  & 4 vecs & $  {3 } $ & $ 4 - 6 {\a}$ & $ {3}$ \\

$4$ & ${\cal N}=4 \times {\cal N}=0|_{{\bf Z}_2\times {\bf Z}_2} $ & 2 vecs & $ {2} $ & $  4 - 4 {\a}$ & $ {2}$ 
 \\

$4$ & ${\cal N}=2|_{{\bf Z}_2} \times {\cal N}=2|_{{\bf Z}_2}  $ & 2 vecs & $  {2} $ & $   4 - 4 {\a}$ & $  {2}$ 
 \\

$4$ &  ${\cal N}=4 \times {\cal N}=0|_{{\bf Z}_4}  $ & pure &$  {1} $ & $  4 - 2 {\a}$ 
& $ {1}$  \\[3pt]

\hline

$3$ & ${\cal N}=2|_{{\bf Z}_2} \times {\cal N}=1|_{{\bf Z}_3}  $ & 1 vec & $  {3} $ & $  {5} - 6 {\a}$ 
& $  {3}$  \\[3pt]

\hline

$2$ & ${\cal N}=2|_{{\bf Z}_2} \times {\cal N}=0|_{{\bf Z}_2}$  & 7 vecs & $   {20 \over 3}- 8 {\a} $ & $  6 - 16 {\a} + 16 {\a^2} $
 & $  {20 \over 3} - {4 \a}$ \\

$2$ & ${\cal N}=2|_{{\bf Z}_2} \times {\cal N}=0|_{\tilde {\bf Z}_4}$  & 5 vecs & $   {6 }- {6 } {\a} $ & $  6 - 14 {\a} + 12 {\a^2} $
 & $  {6 } - {3}{\a}$ \\

$2$ & ${\cal N}=2|_{{\bf Z}_2} \times {\cal N}=0|_{{\bf Z}_2\times{\bf Z}_2} $ & 3 vecs & $  {16 \over 3}-  4{\a} $ & 
$  6 - 12 {\a} + 8 {\a^2} $ 
 & $   {16 \over 3} -{2} {\a}$\\

$2$ & ${\cal N}=2|_{{\bf Z}_2} \times {\cal N}=0|_{{\bf Z}_4}  $ & 1 vec &  $  {14 \over 3}- { 2} {\a} $ & 
$  6 - 10 {\a} + 4 {\a^2} $ &  
$  {14 \over 3} - {\a}$ \\

$2$ & ${\cal N}=1|_{{\bf Z}_3} \times {\cal N}=1|_{{\bf Z}_3}  $ & 1 hyper &$  {9 \over 2} $ & 
$  6 - {9 } {\a} $ & $  {9 \over 2}$ \\[3pt]

\hline

$1$ & ${\cal N}=1|_{{\bf Z}_3} \times {\cal N}=0|_{{\bf Z}_2}  $ &6 vecs, 1 chiral & $   8- {12 } {\a} $ & 
$  {7 } - 20 {\a} + {24} {\a^2} $
 & $  {8} - {6} {\a}$ \\

$1$ & ${\cal N}=1|_{{\bf Z}_3} \times {\cal N}=0|_{\tilde {\bf Z}_4}  $ &4 vecs, 1 chiral& 
$   {15\over 2}- {9 } {\a} $ & 
$  {7 } - 18 {\a} + 18 {\a^2} $
 & $  {15 \over 2} - {9 \over 2} {\a}$\\

$1$ & ${\cal N}=1|_{{\bf Z}_3} \times {\cal N}=0|_{{\bf Z}_2\times{\bf Z}_2}  $ &2 vecs, 1 chiral& 
$   {7}- {6} {\a} $ & 
$  {7 } - 16 {\a} + 12 {\a^2} $
 & $  {7 } - {3 } {\a}$  \\

$1$ & ${\cal N}=1|_{{\bf Z}_3} \times {\cal N}=0|_{{\bf Z}_4}  $ & 1 chiral & $   {13 \over 2}- {3 } {\a} $ & 
$  {7 } - 14 {\a} + 6 {\a^2} $
 & $  {13 \over 2} - {3 \over 2} {\a}$ \\[3pt]
\hline

\end{tabular} }

\caption{ \small Four-gravitons amplitudes at one loop for factorizable orbifolds. 
The polynomials ${\cal P}_1,\, {\cal P}_2$ and ${\cal Q}$
depend on the ratio of Mandelstam variables $\a = {t u / s^2}$.\label{tabAmpgrav}
}

\end{center}
\end{table}

The one-loop amplitudes for all the orbifold theories in Tables~\ref{taborb} and \ref{taborb_moduli} 
and with ${\cal N}=1,2,3,4$ supersymmetry have the following general form,
\begin{multline}
{\cal M}_4^{(1)} =  
{\cal M}_{4,{\cal N}=8}^{(1)}+  i \Big({\kappa \over 2}\Big)^4 {M}_{\mathrm{tree}}  {t u \over s} \left( {  u-t \over s}  {\cal P}_1 
\Big( {t u \over s^2}\Big) \big( I_2(t) - I_2(u) \big)  \right.  \\
+\epsilon  \ {\cal Q} \Big( {t u \over s^2}\Big) \big( I_2(t) + I_2(u) \big)  
  + \left. \ {\cal P}_2 \Big( {t u \over s^2}\Big) \ \Big( t I_3(t) + u I_3(u) - {t u \over 2} I_4(t,u) \Big) \right)  +   {\cal O}(\epsilon)  , 
 \label{ampgrav} 
 \end{multline}
where ${\cal P}_1, {\cal P}_2$ and ${\cal Q}$ are polynomials 
which depend on the particular orbifold theory  and  
${\cal M}_{4,{\cal N}=8}^{(1)}  = i  \left({\kappa \over 2}\right)^4 s t u {M}_{\mathrm{tree}} \big( I_4(s,t) + I_4(s,u) +I_4(t,u)  \big)  $ is the known~\cite{Green:1982sw} one-loop four-graviton amplitude  in the ${\cal N}=8$ theory. All scalar integrals in eqn.~\ref{ampgrav} are to be evaluated in $D=4-2 \epsilon$ dimensions.   The combination of triangle and box integrals in the parenthesis on its second line may be recognized as a multiple of the six-dimensional box integral $ t I_3(t) + u I_3(u) - {t u \over 2} I_4(t,u)  = s I_4^{D=6-2 \epsilon}(t,u)$.

The polynomials ${\cal P}_1, {\cal P}_2$ and ${\cal Q}$ for the 
factorizable orbifold supergravity 
theories from Table~\ref{taborb_moduli} are listed in Table~\ref{tabAmpgrav}.
{Amplitudes} for   ${\cal N}=4$ {theories} have already been studied extensively in 
\cite{Bern:2011rj,Bern:2012cd,Bern:2012gh,Dunbar:1994bn,Dunbar:1999nj}, 
and are included here only for completeness.
We note here that the ${\cal N}=4$, ${\bf Z}_2 \times {\bf Z}_2$ amplitude on row 3 differs 
from the  ${\cal N}=2$, ${\bf Z}_2 \times {\bf Z}_2$ 
amplitude on row 7, despite the theories forming a twin pair (i.e. having the same bosonic content).

Since all matter fields couple canonically to the gravity multiplet through their energy-momentum tensors, all one-loop amplitudes depend 
only on the number of supersymmetries and matter multiplets, and are thus insensitive to the precise nature of the matter couplings.   We can decompose the field content of the theories in terms of $\cN =1$ vector and chiral multiplets, and present unified expressions for all of the amplitudes described  in Table~\ref{tabAmpgrav} 
\bea  
 {\cal P}_1(\alpha)  &=&   \Big( {17 \over 2} - {25 \over 12} \cN + {1 \over 4 }n_V +{1 \over 12 }n_C \Big) - \Big( 6 - {5 \over 2} \cN + {3 \over 2 } n_V - {1 \over 2 } n_C\Big) \alpha \ , \\
 {\cal P}_2(\alpha) &=&   \big( 8 - \cN \big) - \Big( {19 } - {5 } \cN + {n_V } \Big) \a + 2 \Big( 6 - {5 \over 2} \cN + {3 \over 2 } n_V - {1 \over 2 }n_C \Big) \alpha^2  \ , \\
 {\cal Q}(\alpha) &=&   \Big( {17 \over 2} - {25 \over 12} \cN + {1 \over 4 }n_V +{1 \over 12 }n_C \Big) - {1 \over 2}\Big( 6 - {5 \over 2} \cN + {3 \over 2 } n_V - {1 \over 2 }n_C \Big) \alpha \ ,
 \eea
where $n_V$ and $n_C$ are the numbers of vector and chiral multiplets.   We will see in the next subsection that matter amplitudes do not in general 
share this structure and depend on the specific nature of the matter couplings.   

It is interesting to note that the constant term in ${\cal P}_2$ 
does not depend on the matter content of the theory and is always equal to 
${8- \cN }$.  The various multiplets of the existing supersymmetry algebra contribute independently to one-loop amplitudes.
Thus, the terms depending on the polynomials ${\cal P}_1$, ${\cal P}_2$ and ${\cal Q}$ can be read as 
removing from the $\NeqEight$ amplitude the  contribution of the multiplets  that have been orbifolded away. 
See for example the constant term in ${\cal P}_2$.  The form ${8- \cN }$ represents the (ultimately removed) contribution of the $(8-{\cal N})$ 
gravitino multiplets 
that vary under the action of the orbifold group $\Gamma$.
Moreover, one can note that the highest-degree terms in the three polynomials are proportional to each other and the constant terms in ${\cal P}_1$ and ${\cal Q}$  are identical.  

An interesting example is provided by the two apparently different realizations of $\NeqFour$ supergravity coupled 
to two vector multiplets, realized from the double-copy perspective as different ${\bf{Z}_2\times {\bf Z}_2}$ 
orbifolds, which appear on rows  five and six of Tables~ \ref{taborb} and \ref{taborb_moduli}. Their amplitudes turn out to be identical --
\bea 
{{\cal M}^{(1)}_{4,{\bf{Z}_2\times {\bf Z}_2}}} &=& {\cal M}_{4,{\cal N}=8}^{(1)}
+  i \Big({\kappa \over 2}\Big)^4 { M}_{\textrm{tree}} \left( 2 t u { u-t\over s^2}  \big( I_2 (t) - I_2(u) \big) 
+  4  \epsilon  {t u \over s } \big(  I_2(u) +  I_2(t) \big)\no \right. \\
&&  \left. + 4  {tu\over s} \Big(1- {t u \over s^2} \Big) \Big( t I_3 (t) + u I_3(u) - {t u \over 2} I_4(t,u) \Big) \right)
 \ . 
 \label{Neq4w2vectors}
 \eea
One may convince oneself that this is indeed so by noticing that the supersums that govern the four-dimensional cuts 
of all four-point one-loop amplitudes are identical:
\bea && \! \! \! \!\! \! \! 
\Big( \langle 1 l_1 \rangle^4 \langle 2 l_2 \rangle^4 - 2 \langle 1 l_1 \rangle^3 \langle 1 l_2 \rangle \langle 2 l_1 \rangle \langle 2 l_2 \rangle^3 +
 2 \langle 1 l_1 \rangle^2 \langle 1 l_2 \rangle^2 \langle 2 l_1 \rangle^2 \langle 2 l_2 \rangle^2 - 2 \langle 1 l_1 \rangle \langle 1 l_2 \rangle^3 \langle 2 l_1 \rangle^3 \langle 2 l_2 \rangle 
 \no \\  
&&  + \langle 1 l_2 \rangle^4 \langle 2 l_1 \rangle^4 \Big)^2 =  \Big( \langle 1 l_1 \rangle \langle 2 l_2 \rangle -  
\langle 1 l_2 \rangle \langle 2 l_1 \rangle \Big)^4 \Big( 
  \langle 1 l_1 \rangle^2 \langle 2 l_2 \rangle^2 +  \langle 1 l_2 \rangle^2 \langle 2 l_1 \rangle^2 \Big)^2 \ .
\eea
The expression on the left-hand side is the supersum for the 
${\cal N}=2|_{{\bf Z}_2} \times {\cal N}=2|_{{\bf Z}_2}$ theory, 
while the expression on the right-hand side is the supersum for the 
${\cal N}=4 \times {\cal N}=0 |_{{\bf Z}_2\times {\bf Z}_2}$ theory. 
One may also understand this equivalence as a consequence of the fact that,
up to conjugation by $SU(8)$ elements, there is a unique embedding  
of the orbifold group $\Gamma={\bf{Z}_2\times {\bf Z}_2}$ inside the $SU(8)$ 
R-symmetry group of $\NeqEight$ supergravity.
As we will see in the next subsections, matter amplitudes exhibit the same feature.
We will comment in sections \ref{subsecrational} and \ref{subsecdoublecopies}
on the equivalence of the two theories at the level of rational terms and for higher-loop and 
higher-point amplitudes.

A thorough analysis of the non-factorizable orbifolds such as those shown 
in Table~\ref{taborb2} is beyond the scope of this paper. 
However, as  examples, we present in Table~\ref{tabAmpgrav2} the four-graviton amplitudes 
for the pure ${\cal N}=1,2,3$  supergravities. 
We notice the close similarity of the graviton amplitude in the pure ${\cal N}=3$ and ${\cal N}=4$ supergravities (cf. first row of 
Table~\ref{tabAmpgrav2} and fifth row of Table~\ref{tabAmpgrav}). It would be interesting to check whether this similarity 
continues at higher loops.

\begin{table}[t]
\small
\begin{centering}

{ \renewcommand{\arraystretch}{1.4}

\begin{tabular}{|c|ccccc|}
\hline
{\bf $\cN$}& {\bf Orbifold} & {\bf Matter} & $   {\cal P}_1 ( {\a}) $  
& $  {\cal P}_2( {\a})$ &  $  {\cal Q} ( {\a})$\\[3pt]
\hline 

$3$ & ${\bf Z}_5$ & pure  &$   { 5 \over  2} $ & $  {  5 } - {   5} { \a}$ 
 & $  {5 \over 2}$ 
 \\[3pt]

\hline

$2$ & ${\bf Z}_6 $ & pure & $   {  13 \over   3}-  {\a} $ & $   6 - {  9 } {\a} + 2 {\a^2} $ 
 & $  {13 \over 3} - {1 \over 2} {\a}$ 
\\[3pt]

\hline

$1$ & ${\bf Z}_7  $ & pure & $   {77 \over 12}- {7 \over 2 } {\a} $ & 
$  {7 } - 14 {\a} + {7 } {\a^2} $
& $  {77 \over 12} - {7 \over 4} {\a}$ \\[3pt]

\hline

\end{tabular}

}
\caption{ \small Four-gravitons amplitudes at one loop for some non-factorizable orbifolds. \label{tabAmpgrav2}}

\end{centering}

\end{table}

\subsection{Matter amplitudes \label{subsecmat}}

\begin{table}[t]
\small
\begin{center}
 
{ \renewcommand{\arraystretch}{1.4}

\begin{tabular}{|c|cccccc|}
\hline
{\bf $\cN$}& {\bf Orbifold} & {\bf Matter} & $c_1$ & $c_2$ & $c_3$  & $  {\cal P} (\a)$ \\[2pt] 
\hline

$4$ & ${\cal N}=4  \times  {\cal N}=0|_{{\bf Z}_2}  $  & 6 vecs & $ 4 $ & $0$ & $ 0 $  & $4$ \\

$4$ & ${\cal N}=4 \times {\cal N}=0|_{\tilde {\bf Z}_4} $ & 4 vecs & $  {3 } $ & $0$ & $ 0 $ & $   4 $ \\[2pt]

$4$ & ${\cal N}=4 \times {\cal N}=0|_{{\bf Z}_2\times {\bf Z}_2} $ & 2 vecs & $  { 2} $ & $0$ & $ 0 $ & $   4 $ \\[2pt]

$4$ & ${\cal N}=2|_{{\bf Z}_2} \times {\cal N}=2|_{{\bf Z}_2}  $ & 2 vecs & $  { 2} $ & $0$ & $ 0 $ & $   4 $ \\[3pt]

\hline

$2$ & ${\cal N}=2|_{{\bf Z}_2} \times {\cal N}=0|_{{\bf Z}_2}  $ & 7 vecs & $   {14 \over 3} $ & $   {2} $ & $   {16 \over 9} $&
$  6 - 4 {\a}  $ \\

$2$ & ${\cal N}=2|_{{\bf Z}_2} \times {\cal N}=0|_{\tilde {\bf Z}_4}  $ & 5 vecs & $   {4} $ & $   { 2} $ & $   {11 \over 6} $&
$  6 - 4 {\a}  $ \\

$2$ & ${\cal N}=2|_{{\bf Z}_2} \times {\cal N}=0|_{{\bf Z}_2\times{\bf Z}_2} $ & 3 vecs & $  {10 \over 3} $ & $   { 2} $
& $   {17 \over 9} $& 
$  6 - 4 {\a} $ \\[3pt]

\hline

$1$ & ${\cal N}=1|_{{\bf Z}_3} \times {\cal N}=0|_{{\bf Z}_2}  $ & 6 vecs, 1 chiral & $  {5 } $ & $   {3 } $
& $   {8 \over 3} $ & 
$  {7} - 6 {\a} $\\

$1$ & ${\cal N}=1|_{{\bf Z}_3} \times {\cal N}=0|_{\tilde {\bf Z}_4}  $ & 4 vecs, 1 chiral & $  {9 \over 2} $ & $   {3 } $
& $   {11 \over 4} $ & 
$  {7 } - 6 {\a} $  \\

$1$ & ${\cal N}=1|_{{\bf Z}_3} \times {\cal N}=0|_{{\bf Z}_2\times{\bf Z}_2}  $ & 2 vecs, 1 chiral & $  {4} $ & $   {3} $
& $   {17 \over 6} $ & 
$  {7} - 6 {\a} $   \\[3pt]
\hline

\end{tabular} }

\caption{ \small One-loop vector amplitudes $F^{123456} F^{78} \rightarrow F^{123456} F^{78}$ 
for factorizable orbifolds.   ${\cal P}$ is a function of the ratio of Mandelstam variables $\a = {t u \over s^2}$ \label{tabmatter1}}

\end{center}
\end{table}

\begin{table}[t]
\small
\begin{center}

{ \renewcommand{\arraystretch}{1.4}

\begin{tabular}{|c|cccccc|}
\hline
{\bf $\cN$}& {\bf Orbifold} & {\bf Matter} & $c_1$ & $c_2$ & $c_3$  & $  {\cal P} (\a)$ \\[0pt] 
\hline

$4$ & ${\cal N}=4 \times {\cal N}=0|_{{\bf Z}_2}  $ & 6 vecs & $ 4 $ & $0$ & $ 0 $  & $4$ \\

$4$ & ${\cal N}=4 \times {\cal N}=0|_{\tilde {\bf Z}_4} $ & 4 vecs & $  {3 } $ & $0$ & $ 0 $ & $   4 $ \\[2pt]

$4$ & ${\cal N}=4 \times {\cal N}=0|_{{\bf Z}_2\times {\bf Z}_2} $ & 2 vecs & $  { 2} $ & $0$ & $ 0 $  & $   4 $ \\[2pt]

$4$ & ${\cal N}=2|_{{\bf Z}_2} \times {\cal N}=2|_{{\bf Z}_2}  $ & 2 vecs & $  { 2} $ & $0$ & $ 0 $  & $   4 $ \\[3pt]

\hline

$3$ & ${\cal N}=2|_{{\bf Z}_2} \times {\cal N}=1|_{{\bf Z}_3}  $ & 1 vec & $  {3 } $ & $0$  & 
$0 $ & $   {5} $ \\[3pt]

\hline

$2$ & ${\cal N}=2|_{{\bf Z}_2} \times {\cal N}=0|_{{\bf Z}_2}  $ & 7 vecs & $   {8 \over 3} $ & $   {4 } $ &  $  {34 \over 9} $ & 

$  6 - 8 {\a}  $ \\

$2$ & ${\cal N}=2|_{{\bf Z}_2} \times {\cal N}=0|_{\tilde {\bf Z}_4}  $ & 5 vecs & $   {9 \over 3} $ & $   {3 } $ & $   {17 \over 6} $&
$  6 - 6 {\a}  $ \\

$2$ & ${\cal N}=2|_{{\bf Z}_2} \times {\cal N}=0|_{{\bf Z}_2\times{\bf Z}_2} $ & 3 vecs & $  {10 \over 3} $ & $   { 2} $
& $   {17 \over 9} $ & 
$  6 - 4 {\a} $ \\

$2$ & ${\cal N}=2|_{{\bf Z}_2} \times {\cal N}=0|_{{\bf Z}_4}  $ & 1 vec & $  {11 \over 3} $ &
$   {1 } $ & $  {17 \over 18} $ & 

$  6 - 2 {\a} $ \\[3pt]

\hline

\end{tabular}

\caption{ \small One-loop vector amplitudes $F^{125678} F^{34} \rightarrow F^{125678} F^{34}$ 
for factorizable orbifolds. \label{tabmatter2}}}
\end{center}
\end{table} 

\begin{table}
\small
\begin{center}

{ \renewcommand{\arraystretch}{1.4}

\begin{tabular}{|c|cccccc|}
\hline
{\bf $\cN$}& {\bf Orbifold} & {\bf Matter} & $c_1$ & $c_2$ & $c_3$ & $  {\cal P}(\a)$ \\[0pt] 
\hline

$2$ & ${\cal N}=1|_{{\bf Z}_3} \times {\cal N}=1|_{{\bf Z}_3}  $ & 1 hyper & $  {9 \over 2} $ & $   {0} $ & $   {0} $
& $  6 $ \\[3pt]

\hline

$1$ & ${\cal N}=1|_{{\bf Z}_3} \times {\cal N}=0|_{{\bf Z}_2}  $ & 6 vecs, 1 chiral & $  {4} $ & 
$   {4} $ & 
$   {11 \over 3} $  & 
$  {7 }- 8 {\a} $  \\

$1$ & ${\cal N}=1|_{{\bf Z}_3} \times {\cal N}=0|_{\tilde {\bf Z}_4}  $ & 4 vecs, 1 chiral & $  {9 \over 2} $ & 
$   {3 } $ & 
$   {11 \over 4} $  & 
$  {7 }- 6 {\a} $ \\

$1$ & ${\cal N}=1|_{{\bf Z}_3} \times {\cal N}=0|_{{\bf Z}_2\times{\bf Z}_2}  $ & 2 vecs, 1 chiral & $  {5} $ & 
$   { 2} $ & 
$   {11 \over 6} $  & 
$  {7 }- 4 {\a} $ \\

$1$ & ${\cal N}=1|_{{\bf Z}_3} \times {\cal N}=0|_{{\bf Z}_4}  $ & 1 chiral & $  {11 \over 2} $ & $   {1 } $ & 
$   {11 \over 12} $ & 
$  {7 } - 2{\a} $ \\[3pt]

\hline

\end{tabular} }

\caption{ \small One-loop scalar amplitudes $\phi^{1234} \phi^{5678} \rightarrow \phi^{1234} \phi^{5678}$ 
for factorizable orbifolds. \label{tabmatter3}}
\end{center}
\end{table}

Let us now study vector and scalar amplitudes of the form 
$F \bar F \rightarrow F \bar F$ or $\phi \bar \phi \rightarrow \phi \bar \phi$  
with all fields in the same matter supermultiplet. 
Specifically, there are three relevant amplitudes, 
\bea 
&& {{\cal M}_V} \big( F^{123456}F^{123456} F^{78} F^{78} \big) \ , \label{matteramp1} \\
 && {{\cal M}_V}' \big( F^{125678} F^{125678} F^{34} F^{34} \big) \ , \label{matteramp2} \\
 && {{\cal M}}_H \big( \phi^{1234} \phi^{1234} \phi^{5678} \phi^{5678} \big) \ . \label{matteramp3}
 \eea
In some particular cases -- namely all the ${\cal N}=4$ orbifolds and the 
${\cal N}=2$, ${\bf Z}_2 \times {\bf Z}_2 \times {\bf Z}_2 $ theory --
the two vector amplitudes (\ref{matteramp1}) and (\ref{matteramp2}) are equivalent. 
In some other cases, one of the two amplitudes does not exist as the corresponding 
fields are projected out by the orbifold projection. 

All amplitudes are constructed using the \Ansatz,
\be 
{\cal M}^{(1)}_{4} =  \Big({\kappa \over 2}\Big)^4 {{ M}_{\textrm{tree}} \over s^{5-{\cal N}}t u} { \int {d^{4-2 \epsilon}l \over (2 \pi)^{4-2 \epsilon}} } \Big( \sum_{n=1}^3 P^{(n)}_{{10-{\cal N}},0} 
\big(s_{ij}\big) {{  \cal I}}_{4,n}^{D=4-2\epsilon}+ \sum_{m=1}^6 Q^{(m)}_{{9-{\cal N}},N_L} 
\big(s_{ij}, \tau_k \big) {  \cal I}^{D=4-2\epsilon}_{3,m}  \Big) 
\label{ansatzN4mat}
\ee 
where as usual all integrals are in $D=4-2 \epsilon$ dimensions. As for the four-graviton one-loop amplitudes, 
the maximum number of loop momenta in the numerators is $N_L=6-{\cal N}$ for $\cN = 2,3,4$ and $N_L=4$ for $\cN=1$.
After integral reduction all these amplitudes take the following form,
\bea 
{\cal M}^{(1)}_{4} &=& {\cal M}_{4,{\cal N}=8}^{(1)}+   i  \Big({\kappa \over 2}\Big)^4 { M}_{\mathrm{tree}}  { t u \over s} \left(  
\big( c_1 + \epsilon \ c_3 \big)  \big( I_2(t) + I_2(u) \big) + c_2 { u - t \over s}  \big( I_2(t) - I_2(u) \big) \right. \no \\
&& \left. +  {\cal P} \Big( {t u \over s^2}\Big) \Big( t I_3(t) + u I_3(u) - {t u \over 2} I_4(t,u) \Big) \right)  + \ {\cal O}(\epsilon)  \ , 
\label{ampmat}
\eea
which depends on the polynomial ${\cal P}$  and on the constants $c_1, c_2, c_3$. 

We start by listing amplitudes ${\cal M}_V$ of the form (\ref{matteramp1}) in Table~\ref{tabmatter1}. 
As expected, the amplitudes with  fields in 
the same matter multiplet are all divergent, i.e. all constants $c_1$ are different from zero.
The divergence of the ${\cal N}=4$ amplitudes is proportional 
to $(2 + n_v)$, where $n_v$ is the number of vector multiplets, in agreement with 
the results of Fischler, Fradkin and Tseytlin \cite{Fischler:1979yk,Fradkin:1983xs} (see also \cite{Bern:2012gh} 
for a recent discussion of UV properties of these theories
from the perspective of the double-copy property of half-maximal supergravities).
The ${\cal N}=3$ and ${\cal N}=2$  orbifolds with one vector multiplet do not have any non-vanishing amplitude of this class 
because the external fields do not survive the ${\bf Z}_2 \times {\bf Z}_3 $ and ${\bf Z}_2 \times {\bf Z}_4 $ projections.

In Table~\ref{tabmatter2} we list the numerical 
coefficients $c_1, c_2, c_3$ and the polynomial ${\cal P}$ for  amplitudes ${{\cal M}_V}'$ 
of the form (\ref{matteramp2}). The $\cN =1$ orbifold supergravities 
do not have amplitudes of this class, because the external fields in (\ref{matteramp2}) 
do not survive the projection.  
For the ${\cal N}=4$ orbifolds, amplitudes in this class are equal to the ones already listed. 

The amplitudes of ${\cal N}=2$ MESGT with three vector multiplets quoted
in Tables~\ref{tabmatter1} and \ref{tabmatter2} coincide. For this theory the U-duality group
is $SO(2,2)\times SU(1,1)$ which is isomorphic to $SU(1,1)^3$ under which the field strengths
of vector fields and their magnetic duals transform in the eight-dimensional representation $(1/2,1/2,1/2)$.  
Because of the triality symmetry permuting the three factors of $SU(1,1)$ all vectors are on
equal footing with respect to the full U-duality group.

In the case of the ${\cal N}=2$ orbifolds with seven and five vector multiplets we find that ${{\cal M}_V}'$ is
different from ${{\cal M}_V}$.
To explain this,  we note that  the   ${\cal N}=2$ MESGTs with $(n+1)$ vector multiplets and  
symmetric scalar manifolds 
\[ 
\frac{SO(n,2)\times SU(1,1)}{SO(n)\times SO(2) \times U(1)} 
\]
 descend from generic Jordan family of five-dimensional MESGTs  with target manifolds~\cite{GST2}
 \[ 
 \frac{SO(n-1,1)\times SO(1,1)}{SO(n-1)} \ .
 \]
 Under $SO(n,2)\times SU(1,1)$ symmetry group $n+2$ vector field strengths and their magnetic duals transform in
 the representation $(n+2,2)$, i.e. they form $(n+2)$
 doublets of $SU(1,1)$. Pure five-dimensional  ${\cal N}=2$ supergravity reduces to  ${\cal N}=2$ supergravity coupled to one vector multiplet with 
 scalar manifold $\frac{SU(1,1)_0}{U(1)}$. Under the isometry group $SU(1,1)_0$ the field strengths of the graviphoton and the single vector field and their 
 magnetic duals transform in the four dimensional "spin"=$3/2$ representation. This $SU(1,1)_0$ symmetry group can be identified with a  
 diagonal subgroup of $SO(1,2)\times SU(1,1)$ subgroup of the isometry group $SO(n,2)\times SU(1,1)$ of generic Jordan family of MESGTs
 \be
 SU(1,1)_0 \times SO(n-1) \subset SU(1,1) \times SO(1,2) \times SO(n-1) \subset SU(1,1) \times SO(n,2) \ . \quad
 \ee
Under the $SU(1,1)_0$ subgroup vector field strengths and their magnetic duals transform as $n$ doublets ("spin"=$1/2$) and one quartet ("spin"=$3/2$). 
This shows that a certain linear combination of the vector fields couples differently than the rest. This gives a physical interpretation to the result that ${\cal M}_V$ and  ${{\cal M}_V}'$  are different in the ${\cal N}=2$ orbifold supergravities 
with seven and five vector multiplets on rows nine and ten of Table~\ref{taborb_moduli}.

Finally, scalar amplitudes of the form (\ref{matteramp3}) in theories with 
hypermultiplets or chiral multiplets are shown in Table~\ref{tabmatter3}.


As in the case of the one-loop four-graviton amplitudes, the numerical coefficients of the expressions in 
Tables~\ref{tabmatter1}-\ref{tabmatter3}
are not all independent. In particular, the constant term in ${\cal P}$ only depends on the amount of supersymmetry and is equal to
$(8- \cN)$. Moreover, the linear term in ${\cal P}$ is equal to $2 c_2$. Finally, $c_1+c_2$ is always equal 
to the constant term in ${\cal P}_1$ shown in Table~\ref{tabAmpgrav}.

\subsection{Two presentations of $\NeqFour$ supergravity with two vector multiplets \label{subsecrational} }

In previous sections, using unitarity, we calculated many four-point one-loop amplitudes in various matter-coupled  supergravity theories that can be obtained by orbifolding $\NeqEight$ supergravity.  We have also observed that
such amplitudes in the theories on rows 
5 and 6 of Table~\ref{taborb_moduli},  ${\cal N}=2|_{{\bf Z}_2}\times {\cal N}=2|_{{\bf Z}_2}$ 
and ${\cal N}=4\times {\cal N}=0|_{{\bf Z}_2\times {\bf Z}_2}$, are the same in four dimensions. We will demonstrate 
here that  this equality  extends to $D$-dimensions.

Unitarity carried out strictly in four-dimensions may miss non-trivial  contributions known as {\em rational terms}.  
These arise from tree-level amplitudes that vanish identically in four dimensions but 
are non-vanishing in $D\ne 4$ dimensions, such as the all-plus amplitudes in pure Yang-Mills (YM) theory.  So, strictly speaking, in order to verify that one has not missed any contributions, 
it is necessary to use unitarity with  $D$-dimensional constituent amplitudes.
For theories that can be consistently expressed as a dimensional-reduction of a higher-dimensional theory one can simply 
use a higher-dimensional formalism, such as the six-dimensional helicity formalism introduced in
 \cite{Cheung:2009dc, Dennen:2009vk} and used in \cite{Bern:2010qa, Brandhuber:2010mm} for 
 $\NeqFour$ sYM as well as pure YM calculations.

The minimal amount of supersymmetry for a six-dimensional sYM theory 
is ${\cal N}=(1,0)$ or  ${\cal N}=(0,1)$; it reduces to ${\cal N}=2$ sYM theory in four dimensions. 
Moreover, six-dimensional non-supersymmetric YM theory may be realized as an orbifold projection 
of this theory and reduces upon dimensional reduction to pure YM theory coupled to two real scalar fields. 
Thus, from the supergravity theories listed in Table~\ref{taborb_moduli}, only those that have such factors  
should be treated in this formalism when allowing {\em all} states to run across all loops. We evaluate here the various supersums appearing in the corresponding  supergravity calculations and discuss their implications. 
In our discussion we will also label the external states  using a six-dimensional notation; thus, all possible 
four-point amplitudes allowed by supersymmetry are covered simultaneously.
One should note that it is certainly possible to consistently handle the other theories using these same methods, but one must constrain the state-sums to only allow states corresponding to the appropriate theory.  In principle this is just a matter of book-keeping.  For example, to consider pure YM in four-dimensions, using higher-dimensional methods, one simply restricts~\cite{Bern:2010qa} to state-sums that only allow two gluonic states\footnote{Or  $2-\epsilon$ gluonic states depending on the regularization scheme.} to run around the loop.   

The tree-level color-ordered four-point superamplitude of the $(1,1)$ sYM theory in six dimensions is \cite{Dennen:2009vk}:
\be
A_4^\text{tree}(1,2,3,4)=\frac{1}{st}\delta^6(\sum_{i=1}^4 p_i)\delta^4(\sum_{i=1}^4 q_i^A)
\delta^4(\sum_{i=1}^4 {\tilde q}_i^A) \ ,
\label{fourpt6d}
\ee
where 
\bea
p^{AB} = \epsilon^{ab} \lambda^A_a\lambda^B_b
~,~\;
p_{AB}=\frac{1}{2}\epsilon_{ABCD}p^{CD} = {\tilde\lambda}_A^{\dot a}{\tilde\lambda}_B^{\dot b}
\epsilon_{{\dot a}{\dot b}}
%
~,~\;
q_i^A = \lambda^A_{i,a}\eta_{+,i}^a
~,~\;
{\tilde q}_{iA} = {\tilde \lambda}_{i,A}^{\dot a}{\tilde \eta}_{+,i,{\dot a}} \ .~~
\eea
Similarly to the four-dimensional on-shell superspace, component amplitudes are obtained by extracting 
from (\ref{fourpt6d}) the appropriate combinations of Grassmann variables $\eta$.
The six-dimensional on-shell superfield that contains the various component fields is \cite{Huang:2011um}:
\bea
\Phi^{D=6}(\eta, {\tilde\eta}) &=& \phi+\chi^a\eta_a+{\tilde\chi}_{\dot a}{\tilde\eta}^{\dot a}
+\phi'(\eta)^2+g^a{}_{\dot a}\eta_a{\tilde\eta}^{\dot a}
\cr
&&\qquad\qquad
+\phi''({\tilde\eta})^2 + 
{\tilde\lambda}_{\dot a}{\tilde\eta}^{\dot a}(\eta)^2 +{\lambda}^{a}{\eta}_{a}({\tilde \eta})^2
+\phi'''(\eta)^2({\tilde \eta})^2 \ .
\label{6dsuperfield}
\eea
To extract component amplitudes we simply multiply (\ref{fourpt6d}) by the desired superfields (each of which 
contains a single nonzero component field) and integrate over all  Grassmann variables.

In the same way as pure $\NeqTwo$ sYM theory in four dimensions may be realized as an orbifold projection of 
$\NeqFour$ sYM theory,  the ${\cal N}=(1,0)$ sYM theory may  be realized as an orbifold projection
of the  ${\cal N}=(1,1)$ sYM theory: one projects out  the fields $\phi,\phi',\phi'', \phi'''$, ${\tilde \chi}_{\dot a}$ 
and ${\tilde \lambda}_{\dot a}$. All surviving states have exactly one $\eta$ Grassmann variable and up to two ${\tilde\eta}$ Grassmann variables. The ${\cal N}=(0,1)$ sYM theory is obtained by conjugation.

Each of the supergravity tree-level amplitudes 
can be expressed as a sum over products of tree-level color-ordered (s)YM amplitudes using the KLT  relations~\cite{KLT}.  As discussed at length in \cite{Bern:1998ug}, cuts of  such supergravity theories factorize over these relations to sums over products of color-ordered cuts of  the corresponding (s)YM amplitudes.   
The four (s)YM cuts which determine 
the cuts of all one-loop four-point amplitudes of supergravity theories we analyze here  can be easily computed though the appropriate restriction on the dependence of the Grassmann 
$\delta$-functions in (\ref{fourpt6d}) on $\eta$ and ${\tilde\eta}$. They are:
\bea
{\cal S}^{(1,1)}_4&=&\sum_s A_{4;(1,1)}^\text{tree}(1,2,l^s_1,l^s_2)A_{4;(1,1)}^\text{tree}(l^s_2,l^s_1,3,4) 
 = \frac{\delta^4(q_\text{ext})\delta^4({\tilde q}_\text{ext})}{(k_2+l_1)^2(k_3-l_1)^2} \ ,
\label{11theory}
\\
{\cal S}^{(0,0)}_4&=&\sum_s A_{4;(0,0)}^\text{tree}(1,2,l^s_1,l^s_2)A_{4;(0,0)}^\text{tree}(l^s_2,l^s_1,3,4) =\frac{1}{s^2_{12}\,(k_2+l_1)^2(k_3-l_1)^2}
\cr
&&\qquad\qquad\qquad\quad\,
\times\epsilon_{ABCD}q_1{}^Aq_2{}^B
~\epsilon_{EFGH}q_3{}^Eq_4{}^F(l_1)^{CG}(l_2)^{DH}
\cr
&&\qquad\qquad\qquad\qquad
\epsilon^{A'B'C'D'}{\tilde q}_{1,A}{\tilde q}_{2,B'}
~\epsilon^{E'F'G'H'}{\tilde q}_{3,E}{\tilde q}_{4,F'}
(l_1)_{C'G'}(l_2)_{D'H'} \ ,~~
\label{00theory}
\\
{\cal S}^{(1,0)}_4&=&\sum_s A_{4;(1,0)}^\text{tree}(1,2,l^s_1,l^s_2)A_{4;(1,0)}^\text{tree}(l^s_2,l^s_1,3,4) 
=\frac{\delta^4(q_\text{ext})(2 l_1\cdot l_2)}{s_{12}^2(k_2+l_1)^2(k_3-l_1)^2}
\cr
&&\qquad\qquad\qquad\quad\,
\times\epsilon^{A'B'C'D'}{\tilde q}_{1,A}{\tilde q}_{2,B'}
~\epsilon^{E'F'G'H'}{\tilde q}_{3,E}{\tilde q}_{4,F'}
(l_1)_{C'G'}(l_2)_{D'H'} \ ,~~
\label{10theory}
\\
{\cal S}^{(0,1)}_4&=&\sum_s A_{4;(0,1)}^\text{tree}(1,2,l^s_1,l^s_2)A_{4;(0,1)}^\text{tree}(l^s_2,l^s_1,3,4) =
\frac{\delta^4({\tilde q}_\text{ext})(2 l_1\cdot l_2)}{s_{12}^2(k_2+l_1)^2(k_3-l_1)^2}
\cr
&&\qquad\qquad\qquad\quad\,
\times \epsilon_{ABCD}q_1{}^Aq_2{}^B
~\epsilon_{EFGH}q_3{}^Eq_4{}^F(l_1)^{CG}(l_2)^{DH} \ .
\label{01theory}
\eea
These unitarity cuts are all that is required to construct the $D$-dimensional version of all the one-loop four-point 
amplitudes of the supergravities listed on rows  one, five, six and eleven of Table~\ref{taborb_moduli}. 

While the resulting expressions are slightly more involved, the construction described above can be 
modified to also accommodate the pure YM theory coupled to other even numbers of scalar fields. 
For example, for a vector field coupled with six real scalars in four dimensions
(the ${\cal N}=0|_{{\bf Z}_2}$ theory) it is necessary 
to project onto the invariant part of the ${\bf Z}_2$ transformation 
\be
{\bf Z}_2: (\eta, {\tilde\eta}) \mapsto (-\eta, -{\tilde\eta}) \ ;
\ee
This transformation projects out all fermions in eq.~(\ref{6dsuperfield}) while keeping all bosonic fields.

Using eqs.~(\ref{11theory})-(\ref{01theory}) we can see that the observed equality of the four-point 
one-loop amplitudes of the ${\cal N}=2|_{{\bf Z}_2}\times {\cal N}=2|_{{\bf Z}_2}$ and 
${\cal N}=4\times {\cal N}=0|_{{\bf Z}_2\times {\bf Z}_2}$ theories holds in higher dimensions.
Indeed, we note that the four supersums (\ref{11theory})-(\ref{01theory}) are related by
\be
{\cal S}^{(1,1)}_4{\cal S}^{(0,0)}_4={\cal S}^{(0,1)}_4{\cal S}^{(1,0)}_4 \ ;
\ee
This implies that all cuts of the one-loop four-point amplitudes of the theories on rows five and six of 
Table~\ref{taborb_moduli} -- and therefore also their complete one-loop four-point amplitudes -- are the same in any $D\le 6$.
This includes the ``rational'' one-loop amplitudes in the  ${\cal N}=2|_{{\bf Z}_2}\times {\cal N}=2|_{{\bf Z}_2}$ supergravity theory whose integrand vanishes identically on four dimensional cuts.  This might come as a surprise since no such amplitude contributes to either of the sYM factors.  How could this be reconciled with the double-copy property of these theories?  They are generated by terms that  integrate to zero in four dimensions in the two gauge theory factors, but whose double-copy no longer vanishes under integration.

\renewcommand{\theequation}{4.\arabic{equation}}
\setcounter{equation}{0}

\section{$\NeqOne$ and $\NeqTwo$ sYM one-loop amplitudes from color/kinematics duality}

Explicit calculations in the $\NeqFour$ sYM theory revealed that the integrands of scattering amplitudes 
in this theory may be organized such that they exhibit a certain duality between color and kinematics.
Such a duality, which we will review below, is believed to hold in all theories whose amplitudes may 
be organized in terms of cubic graphs. 

\subsection{On the color/kinematics duality and the double-copy structure of supergravity amplitudes}

The general expression of the  dimensionally-regularized $L$-loop 
$m$-point scattering amplitude in such a theory is 
\begin{equation}
{\cal A}^{L-\text{loop}}_m\ =\ 
i^L \, g^{m-2 +2L } \,
\sum_{i\in \Gamma}{\int{\prod_{l = 1}^L \frac{d^D p_l}{(2 \pi)^D}
\frac{1}{S_i} \frac {n_i C_i}{\prod_{\alpha_i}{p^2_{\alpha_i}}}}}\,, 
\label{LoopGauge} 
\end{equation}
where $g$ is the gauge coupling constant.  The sum runs over the complete set $\Gamma$ of 
$m$-point $L$-loop graphs with only cubic (trivalent) vertices, including all permutations of external 
legs, the integration is over the $L$ independent loop momenta $p_l$  
and the denominator is determined by the product of  all propagators of the corresponding 
graph.
The coefficients $C_i$ are the color factors obtained by assigning to every three-vertex in the 
graph a factor of the structure constant
\be
{\tilde f}^{abc} = i \sqrt{2} f^{abc}=\Tr([T^{a},T^{b}]T^{c})\,,
\ee
while respecting the cyclic ordering of edges at the vertex.
The hermitian generators $T^a$ of the gauge group are normalized via
$\Tr(T^a T^b) = \delta^{ab}$.
The coefficients $n_i$ are kinematic numerator factors depending on
momenta, polarization vectors and spinors. For supersymmetric
amplitudes in an on-shell superspace, they will also contain Grassmann 
parameters.
The symmetry factors $S_i$ of each graph remove any overcount
introduced by summing over all permutations of external legs (included by 
definition in the set $\Gamma$), as well as any internal automorphisms of 
the graph -- i.e. symmetries of the graph with fixed external legs.

The color/kinematics duality~\cite{BCJ} is manifest when the kinematic numerators of a graph representation of the amplitude satisfy antisymmetry and (generalized) Jacobi relations around each propagator -- in one-to-one correspondence with the color-factors.  
That is, schematically for cubic-graph representations, it requires that
\begin{equation}
C_i + C_j + C_k =0 \qquad  \Rightarrow \qquad  n_i + n_j + n_k =0 \, .
\label{BCJDuality}
\end{equation} 
The existence of such representations are conjectured~\cite{BCJLoop} to hold  all loop orders and to all multiplicities. 

There is by now substantial evidence for the color/kinematics duality, especially 
at tree level~\cite{OtherTreeBCJ,virtuousTrees,Square,Oconnell},
where explicit representations of the numerators in terms of partial
amplitudes are known for any number of external legs~\cite{ExplicitForms}.  
A partial Lagrangian understanding of the duality has also been given~\cite{Square}.  
An alternative trace-inspired presentation of the duality relation~(\ref{BCJDuality}),
which emphasizes its group-theoretic structure, was described in~\cite{Trace}. 

Once amplitudes of a gauge theory are organized to exhibit color/kinematics duality~(\ref{BCJDuality}),
numerator factors of $L$-loop $m$-point amplitudes in several supergravity or matter-coupled 
supergravity theories  may be obtained in a straightforward way by simply multiplying together kinematic 
numerator factors of this and other gauge theories~\cite{BCJ,BCJLoop},
\begin{equation}
 {\cal M}^{L-\text{loop}}_m = i^{L+1} \, \left(\frac{\kappa}{2}\right)^{m-2+2L} \,
\sum_{i\in\Gamma} {\int{ \prod_{l = 1}^L \frac{d^D p_l}{(2 \pi)^D}
 \frac{1}{S_i}
   \frac{n_i {\tilde n}_i}{\prod_{\alpha_i}{p^2_{\alpha_i}}}}} \,, 
\hskip .7 cm 
\label{DoubleCopy}
\end{equation}
where $\kappa$ is the gravitational coupling.  The factors $n_i$ are the numerator factors 
of the amplitude satisfying color/kinematics duality (\ref{BCJDuality}) and
${\tilde n}_i$ the numerator factors of the amplitude of a second, potentially different gauge-theory 
(at the same loop order and with the same number of external legs) and the sum runs over the same 
set of graphs as in eq.~(\ref{LoopGauge}); moreover, they are not required to satisfy the duality.
The construction (\ref{DoubleCopy}) is expected to hold in a large class of (super)gravity theories
in which the field content may be written as the tensor product of the field content of two gauge 
theories, such as the factorized orbifold supergravity theories discussed in previous sections.
For $\NeqEight$ supergravity both $n_i$ and  ${\tilde n}_i$ are numerators 
of amplitudes of $\NeqFour$ sYM theory.  

The  KLT relations \cite{KLT} described in earlier sections, emerge from the double-copy  construction at 
tree-level when the Jacobi satisfying cubic tree-graph numerators are expressed in terms of color-ordered partial 
amplitudes~\cite{BCJ} -- the so-called amplitude-encoded representations. 

The double-copy formula (\ref{DoubleCopy}) has been
proven~\cite{Square} for both pure gravity and for $\NeqEight$
supergravity tree amplitudes, under the assumption that the
duality~(\ref{BCJDuality}) holds in the corresponding gauge theories,
pure YM and $\NeqFour$ sYM theory, respectively.  The
nontrivial part of the loop-level conjecture is the existence of a
representation of gauge-theory amplitudes that satisfies the duality
constraints.  The multi-loop double-copy property was explicitly exposed for the
three-loop four-point~\cite{BCJLoop} and the four-loop four-point~\cite{Bern:2012uf} 
amplitudes of $\NeqEight$ supergravity.
The color/kinematics duality and double-copy property have
also been confirmed in one- and two-loop five-point amplitudes in
$\NeqEight$ supergravity~\cite{loop5ptBCJ}.  For less-than-maximal
supergravities, the double-copy property has been checked explicitly
for the one-loop four- and five-graviton amplitudes of ${\cal N} =
4,5,6$ supergravity~\cite{N46Sugra} by showing it reproduces known
results~\cite{DunbarEttle}.  
The  infrared divergences and other properties of the four-graviton 
amplitudes in these theories have also been shown to be consistent
with it at two loops  \cite{N46Sugra2} and to all orders \cite{Oxburgh:2012zr}.

The color/kinematics duality and the double-copy property were also discussed and 
shown to hold under certain conditions in the presence of deformations by 
higher-dimension operators \cite{Broedel:2012rc}.  Additionally the duality and double-copy have 
been shown to lead to some suggestive relations between certain ${\cal N} \ge 4$ supergravity 
and subleading-color sYM amplitudes~\cite{SchnitzerBCJ}.

In  the explicit multi-loop calculations carried out in the maximally supersymmetric sYM theory 
it was possible to choose Jacobi-satisfying kinematic numerator factors $n_i$ which 
respect the symmetries  of graphs; that is, if two graphs are obtained from each other by some map
of their edges labels, then so are their 
kinematic numerator factors. Sign factors in such maps are taken to be the same as the
sign appearing in the corresponding map of color factors. 
 Intriguingly the symmetric double-copy representations of the maximally supersymmetric 
gravity theories through four-loops at four-point and two-loops at five-point
make manifest the ultraviolet-behavior of the theory~\cite{BCJLoop, Bern:2012uf, loop5ptBCJ}.   
The double-copy property has also been recently shown to directly lead to 
certain UV cancellations in half-maximal supergravities~\cite{Bern:2012gh}.

Through the calculation of the four-gluon amplitudes in pure $\NeqOne$ and $\NeqTwo$ 
sYM theories  such external-state symmetry  restrictions should not be expected, and we are 
naturally led to presentations of amplitudes asymmetric in external state labels.  We find that
in this case it is quite possible to find representations that do not make the UV properties manifest
and it is such  representations we present below.

\subsection{Color/kinematics duality in the $\cN=1$ and $\cN=2$ sYM theories}

\begin{figure}[ht]
\begin{center}
\includegraphics[height=30mm]{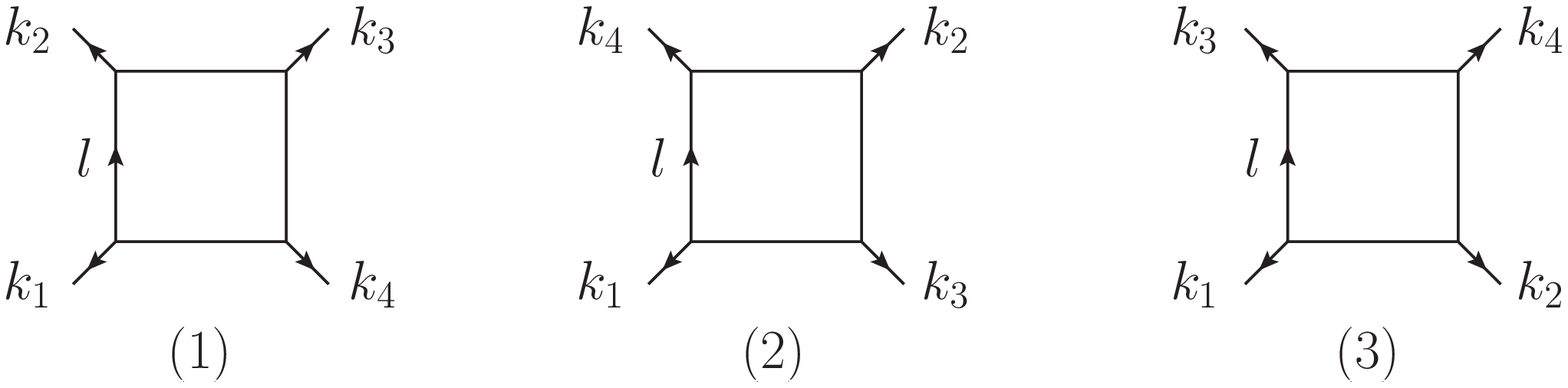}
\caption{ \small Box  graphs contributing to the four-point one-loop amplitudes in ${\cal N}\leq4$ sYM theories.   \label{graphs_p1}}
\end{center}
\end{figure}

\begin{figure}[ht]
\begin{center}
\includegraphics[height=62mm]{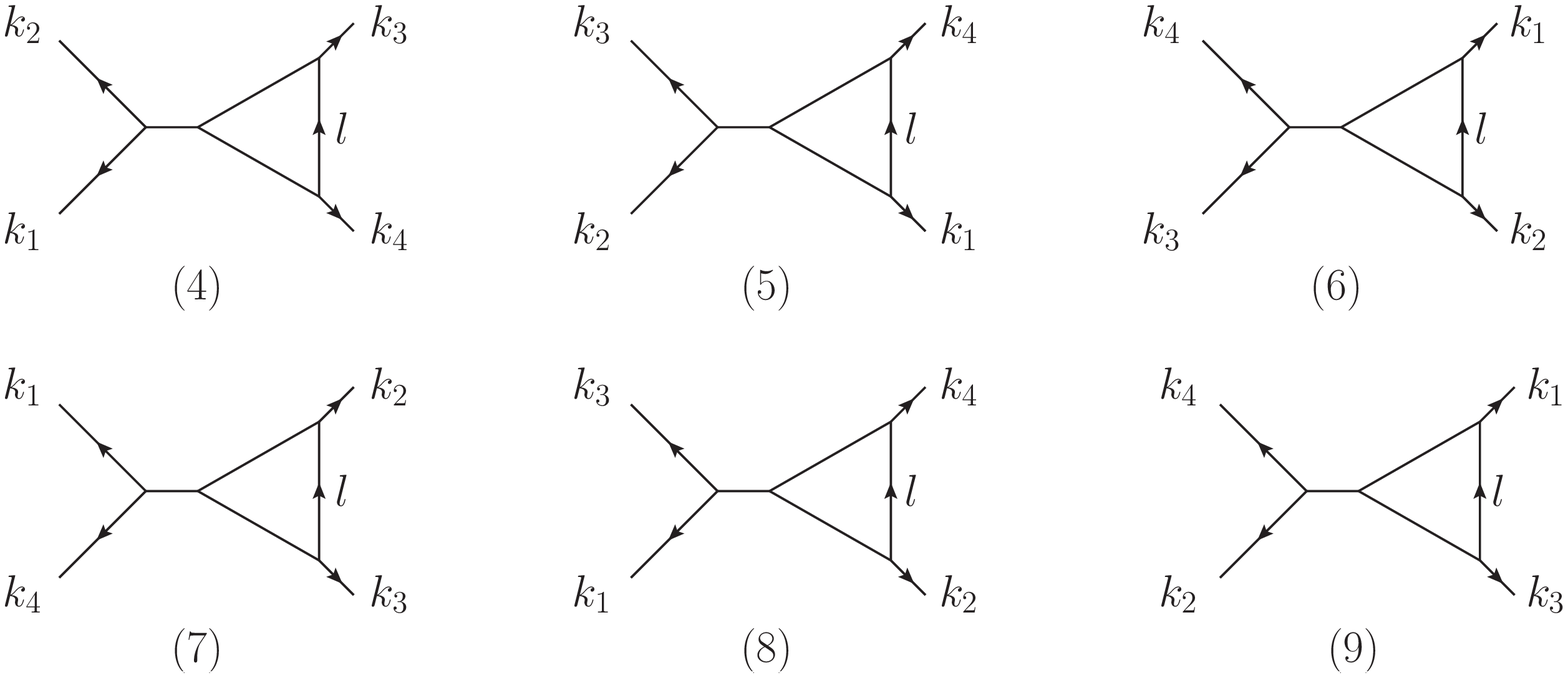}
\caption{ \small Graphs with triangle subgraphs contributing to the  four-point  one-loop amplitudes in ${\cal N}<4$ sYM theories.      \label{graphs_p2}}
\end{center}
\end{figure}

\begin{figure}[ht]
\begin{center}
\includegraphics[height=45mm]{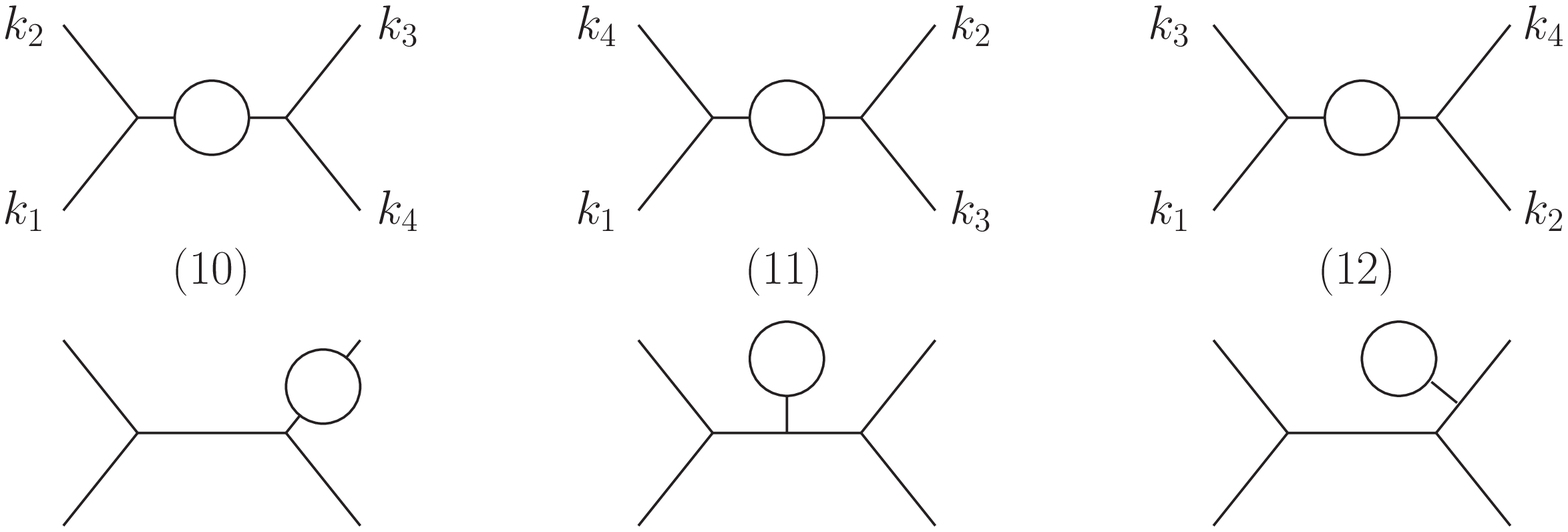}
\caption{ \small Cubic graphs containing bubble and tadpole subgraphs; they need not contribute to the 
four-gluon amplitudes in the $\NeqOne$ and $\NeqTwo$ sYM theory.    \label{nocontrib_graphs}}

\end{center}
\end{figure}

In general, the one-loop four-point amplitude in any gauge theory with fields in the adjoint representation 
is constructed in terms of the  graphs shown in figs.~\ref{graphs_p1} and \ref{graphs_p2}  
as well as graphs containing 
bubble and tadpole subgraphs whose topologies are shown in fig.~\ref{nocontrib_graphs}.
To construct the relevant Jacobi relations we start with the box graphs, iteratively select one of its internal lines and
carry out the Jacobi transformation (\ref{BCJDuality}) of the numerator factors while paying close attention to the 
sign factors arising from reordering the lines at each vertex. For example, the numerator Jacobi relations on the 
edge carrying momentum $l$ in the first graph in fig.~\ref{graphs_p1} and the first graph in fig.~\ref{graphs_p2} are 
\bea
n_1(k_1,k_2,k_3,q+k_2+k_3) -n_3(k_1,k_2,k_3,q+k_3)-n_4(k_1,k_2,k_3,-q) &=& 0
\cr
n_4(k_1,k_2,k_3,k_3+q)+n_4(k_1,k_2,k_3,k_1+k_2+k_3-q)-n_{10}(k_1,k_2,k_3,-q)&=&0 \ ,~~~
\label{egJacobi}
\eea
where $n_{10}$ is the kinematic numerator of the first graph in fig.~\ref{nocontrib_graphs} and $q$ is a generic momentum. 
In general, the Jacobi relations single out 
(non-unique) subsets of color graphs -- master graphs \cite{Carrasco:2011hw, Bern:2012uf} -- whose 
numerators determine all the other ones through Jacobi transformations. 
In our case it is not difficult to see that eqs.~(\ref{egJacobi}), (\ref{fullJacobi})  
allow us to express the numerator factors of graphs with triangle subintegrals in terms of the numerator factors
of the box integrals, the numerator factors of graphs with bubble sub-integrals in terms of those of graphs with 
triangle sub-integrals and so on. In other words, the three box graphs can be chosen as master graphs.

This construction is valid in any gauge theory with adjoint matter at one loop level. 
We found that  in the cases of $\cN=1$ and $\cN=2$ sYM theory  it is also possible to impose a further requirement.    
While  generic one-loop Jacobi-satisfying representations have bubble contributions, for these theories, 
the bubble and tadpole graphs can be chosen to have vanishing numerator factors; in particular, we can require that $n_{10,11,12}(k_1,k_2,k_3,q)=0$.  Such a requirement could have been inconsistent with the cut conditions, but is not for these theories.

While in ${\cal N}=4$ sYM the  numerator 
factors of the three box graphs in fig.~\ref{graphs_p1} can be related by the relabeling of the graphs' external legs, 
for $\NeqOne$ and $\NeqTwo$ theories such a symmetric choice of numerator factors does not appear to be 
possible.  Without loss of generality, we will choose $k_1$ and $k_2$ to be associated with negative helicity gluons, and $k_3$ and $k_4$ to be associated with positive helicity gluons.

To determine the kinematic numerators associated with these three master graphs we proceed to construct an \Ansatz~  for each of them, 
subject to the constraints mentioned above which are  also sufficiently large to solve all the unitarity constraints. 
The known results on the structure of 
${\cal N}=1$ and ${\cal N}=2$ amplitudes \cite{Bern:1995db} suggests that this \Ansatz~  should contain at least two powers 
of the loop momentum. Moreover, the presence of inverse factors of Mandelstam invariants \cite{Bern:1995db} suggests 
that this \Ansatz~  should also contain such factors (and thus is mildly nonlocal).  Last, factors of spinor products are 
necessary to account for the helicity of the scattered particles. We will choose this factor to be the completely symmetric 
combination
\bea
istA(1^{-} \,2^{-}\, 3^{+} \,4^{+}) = \langle 12\rangle^2[34]^2 \ .
\eea
such a choice may in principle introduce additional $s^{-1}$ and $t^{-1}$ factors compared to the expected 
dependence of amplitudes on Mandelstam invariants. The inclusion of inverse factors of Mandelstam invariants 
in the kinematic numerator factors, organized in particular as inverse powers of Gram determinants, was  shown 
\cite{Broedel:2011pd} to be necessary already at tree-level for color-kinematic satisfying representations if one 
requires that the external state information is encoded symmetrically by a tree-level amplitude factor.

A further departure form the structure of the $\NeqFour$ sYM integrand is the presence of parity-odd terms. While 
such terms necessarily integrate to zero, they are required for the cut conditions to be satisfied. They are also crucial 
for the double-copy construction of gravity amplitudes\footnote{Nontrivial contribution to supergravity amplitudes from 
terms that vanish upon integration in gauge theory was noticed in $\NeqEight$ supergravity at four loops 
in \cite{Bern:2012uf}.}. 
Collecting everything, the \Ansatz~  for the kinematic numerator factors of the master graphs is 
\bea
\label{ni}
n_i &=& stA(k_1,k_2,k_3,k_4)\,N_i
\\
N_i &=&\frac{1}{(stu)^2}\Big(P_{i; (6,2)}(\tau_{l,k_1},\tau_{l,k_2},\tau_{l,k_3};s,t) 
                                                              + 4i \varepsilon(k_1,k_2,k_3,l)  \,P_{i; (4,0)}(\tau_{l,k_1},\tau_{l,k_2},\tau_{l,k_3};s,t)\Big) \ ,
\nonumber
\eea
where  $i=1,2,3$ labels the master graph,   
$P_{i; (j,k)}(\tau_{l,k_1},\tau_{l,k_2},\tau_{l,k_3};s,t)$ are degree-$j$ polynomials which 
also have degree-$k$ in their first three arguments, and $\varepsilon(k_1,k_2,k_3,l)=
\varepsilon_{\mu\nu\rho\sigma}k_1^\mu k_2^\nu k_3^\rho l^\sigma$. We recognize the overall factor as the square of the inverse 
of the Gram determinant of the three independent momenta.   Depending on the helicity configuration it is moreover 
possible to choose two of the three odd terms to vanish identically. Recall that we chose for
particles $1$ and $2$ to have negative helicity, so in this case graphs 1 and 3 will have vanishing parity-odd 
terms. Using the above \Ansatz,  Jacobi constraints and imposing the absence of bubbles and tadpoles leaves us 
with $45$~free parameters which are theory dependent and to be constrained by cuts. Due to the Jacobi relations, 
which determine six numerator factors in terms of those of the three master graphs, this number of free parameters 
is far smaller than the number of free parameters of an \Ansatz~  covering all nine contributing integrals.

\subsection{Four-gluon amplitudes in $\NeqOne$ and $\NeqTwo$ sYM theories}

To determine the remaining $45$ coefficients we require that the cuts of the \Ansatz~  are the same as the 
unitarity cuts of the one-loop four-gluon amplitudes in these theories evaluated directly in terms of state sums over products of tree-level amplitudes.
We could use either color-dressed cuts or, alternatively, project the \Ansatz~  onto its color-ordered and color-stripped 
components and use color-ordered cuts. In either case the supersums (i.e. the sum over intermediate states) are 
evaluated through the general strategy \cite{Bern:2009xq} that relates them to the corresponding $\NeqFour$ supersums:
\begin{multline}
{\cal S}_\NeqFour = (A+B+C+\dots)^4 ~\longrightarrow~\\
{\cal S}_{{\cal N}<4} = (A+B+C+\dots)^{\cal N}(A^{4-{\cal N}}+B^{4-{\cal N}}+C^{4-{\cal N}}+\dots) \ ,
\end{multline}
where $A^4, B^4, $ etc. are the contribution of gluon intermediate states to a given MHV amplitude.  The MHV-vertex expansion straightforwardly carries this organization of amplitudes in terms of ${\cal N}=4$ gluonic contributions to any N${}^{k\ge0}$MHV amplitude as a sum over such expressions~\cite{Bern:2009xq}. 
The presence of supersymmetry implies  \cite{BDDK_cut_constructibility}
that one-loop N${}^{k\ge0}$MHV amplitudes in these theories are four-dimensional cut-constructible, at least at the integrated level, and therefore, as long as the 
amplitudes are used in four dimensions, a four-dimensional cut calculation is sufficient.  
Here we will require that the \Ansatz~  satisfies the six distinct  four-dimensional color-ordered two-particle cuts.  The only guarantee on the supergravity side is that the double-copy amplitude will satisfy all four-dimensional cuts.   Verification on higher-dimensional cuts would be sufficient to demonstrate that there are no missing contributions to the gravity amplitude.
Additional work is necessary to construct ${\cal N}<4$ sYM amplitudes in 
$D$-dimensions. While these (unrenormalized) amplitudes are divergent already in four dimensions \cite{Bern:1995db}, their 
$D$-dimensional expressions are necessary to identify the critical dimension of the supergravity theories that
have them as one of the double-copy factors.

As is well-known, one-loop amplitudes in gauge and supergravity theories may be decomposed into the independent 
contributions from the various fields running in the loop. These contributions may also be organized in multiplets of the 
unbroken symmetries; in particular, for ${\cal N}$-extended pure sYM theories, one-loop $n$-point gluon amplitudes 
may be organized as
\be
A^{(1)}_{n,{\cal N}=4} =A^{(1)}_{n,{\cal N}=4} - (4-{\cal N})A^{(1)}_{n, \text{chiral}}\ ,
\ee
where $A^{(1)}_{n, \text{chiral}}$ is the contribution of a single $\NeqOne$ chiral multiplet in the loop. 
Since the Jacobi relations are linear one should expect that numerator  factors obeying color/kinematics duality should 
have similar properties; that is
\be
\label{Ni}
N^{\cal N}_i=N_i^{{\cal N}=4} - (4-{\cal N})N_i^{\text{chiral}} 
\quad, \quad\quad
N_{1,2,3}^{{\cal N}=4}=1
~,~~
N_{4,5,6,7,8,9}^{{\cal N}=4}=0 \ .
\ee
This is indeed what we find.

The six two-particle cut conditions fix uniquely the parity-odd part of the chiral multiplet contribution as well as 
the parity-even part of two of the nine graphs -- graphs $4$ and $6$ in fig~\ref{graphs_p2}. The remaining factors 
depend on $10$ free parameters; this is a reflection of the generalized-gauge symmetry \cite{BCJ, BCJLoop} that allows the addition of  
zero to an amplitude in the form of any arbitrary function multiplied by the color-space Jacobi identity.  
This representation freedom at the integrand level does not affect the content of the amplitude -- any value of these parameters is consistent with the cuts of the theory, and therefore cannot affect the resulting integrated expressions.   Indeed, upon integral reduction\footnote{In the 
process of reducing to an integral basis we must drop all bubble-on-external-line  integrals as such integrals should not 
appear at this loop order in supersymmetric theories. Here they may not cancel automatically and we did not explicitly
exclude them in order to preserve the simplicity of the solution.} 
to basis integrals all free parameters disappear and we recover the classic results of ref.~\cite{Bern:1995db}. 
To present our results we make a particular choice for the free parameters which simplifies the numerator factors; 
with this choice the numerator factors corresponding to the chiral multiplet contribution is given by:
%
\bea
N^\text{chiral}_1\!\!&=&\!\! \frac{1}{s}(\tau_{l,k_1}-\tau_{l,k_2})
                                    +\frac{2}{stu}(s\tau_{l,k_1}\tau_{l,k_2} 
                                   			 -u\tau_{l,k_1}\tau_{l,k_3} 
			 			-t\tau_{l,k_2}\tau_{l,k_3})
\cr
N^\text{chiral}_2\!\!&=&\frac{1 }{s^2}(t\tau_{l,k_2}+s\tau_{l,k_3}-u\tau_{l,k_1})
 +\frac{2}{stu}(s\tau_{l,k_1}\tau_{l,k_2} 
                        -u\tau_{l,k_1}\tau_{l,k_3} 
                        -t\tau_{l,k_2}\tau_{l,k_3})
 -\frac{2 i }{s^2}\varepsilon_{k_1,k_2,k_3,l}
\cr
N^\text{chiral}_3\!\!&=&-\frac{1  }{u}(\tau_{l,k_3}+\tau_{l,k_1})
+\frac{2}{stu}(s\tau_{l,k_1}\tau_{l,k_2} 
                        -u\tau_{l,k_1}\tau_{l,k_3} 
                        -t\tau_{l,k_2}\tau_{l,k_3})
\cr
N^\text{chiral}_4\!\!&=&\!\!\!
-\frac{1}{u}(\tau_{l,k_1}+\tau_{l,k_3})+\frac{1}{t}(\tau_{l,k_2}+\tau_{l,k_3})
\cr
N^\text{chiral}_5\!\!&=&  \; \, \frac{ t } {s^2 u} (s(\tau_{l,k_1} +\tau_{l,k_3}) - u(\tau_{l,k_1}+\tau_{l,k_2}))
              +\frac{2 i}{s^2}\varepsilon_{k_1,k_2,k_3,l}
\label{chiralmultiplet}              
\\
N^\text{chiral}_6\!\!&=&0
\cr
N^\text{chiral}_7\!\!&=&\frac{1}{s^2}(t\tau_{l,k_1}-u\tau_{l,k_2}-s\tau_{l,k_3})
              -\frac{2 i }{s^2}\varepsilon_{k_1,k_2,k_3,l}
\cr
N^\text{chiral}_8\!\!&=&\; \frac{u}{s^2t}(s(\tau_{l,k_2}+\tau_{l,k_3})-t(\tau_{l,k_1}+\tau_{l,k_2}))
               -\frac{2 i }{s^2}\varepsilon_{k_1,k_2,k_3,l}
\cr
N^\text{chiral}_9\!\!&=&\!\!\!
\frac{1}{s^2}(t\tau_{l,k_1}-u\tau_{l,k_2}+s\tau_{l,k_3})
               -\frac{2 i }{s^2}\varepsilon_{k_1,k_2,k_3,l}
\nn
\eea
Note that, unlike the $\NeqFour$ sYM theory, the chiral multiplet numerator factors 
have both a parity-even and a parity-odd component already for the four-point amplitudes. 
On symmetry ground these terms integrate to zero; they however make nontrivial contributions 
to the corresponding supergravity amplitudes though the double-copy construction (\ref{DoubleCopy})
which we discuss in the next section.   Carrying out the reduction to an integral basis we reproduce 
the classic expressions of \cite{Bern:1995db} through ${\cal O}(\epsilon^0)$.

\subsection{Orbifold supergravities as double-copies \label{subsecdoublecopies}}

Together with the known expressions for the one-loop ${\cal N}=4$ sYM amplitudes which automatically exhibit 
color/kinematics duality, the amplitudes constructed in the previous subsection allow us to recover through the 
double-copy construction
\bea
{\cal A}^{(1)}_4 & =& 
i \, g^{4} \,
\sum_{i=1}^9
{\int {\frac{d^D l}{(2 \pi)^D} \frac {n_i C_i}{\prod_{\alpha_i}{p^2_{\alpha_i}}}}}\,, 
\label{LoopGauge1loop} 
\\
 {\cal M}^{(1)}_4 &=& - \left(\frac{\kappa}{2}\right)^{4} \,
\sum_{i=1}^9 {\int{  \frac{d^D l}{(2 \pi)^D}
   \frac{n_i {\tilde n}_i}{\prod_{\alpha_i}{p^2_{\alpha_i}}}}} \,, 
\hskip .7 cm 
\label{DoubleCopy1loop}
\eea
some of the four-point 
amplitudes found though direct computation in sec.~\ref{direct_calculation} for  the factorized orbifold supergravity 
theories listed in Table~\ref{taborb_moduli}. In eqs.~(\ref{LoopGauge1loop}) and (\ref{DoubleCopy1loop}) the numerator 
factors $n_i$ and ${\tilde n}_i$ are given by eqs.~(\ref{ni}), (\ref{Ni}) and (\ref{chiralmultiplet}).
We will emphasize here the main differences from the ${\cal N}\ge 4$ supergravity theories 
in which one of the two copies is $\NeqFour$ sYM theory and discuss the theories in which both factors 
are the $\NeqOne$ and $\NeqTwo$ sYM theories studied in the previous section. While we will focus on
graviton amplitudes, it is important to stress that, due to the reduced supersymmetry, there exist additional 
four-point one-loop scattering amplitudes not related to these by supersymmetric Ward identities; some of them are evaluated in sec.~\ref{direct_calculation}.
%

The essential feature of the one-loop four-point amplitude of $\NeqFour$ sYM theory is that its kinematic numerator 
factors are independent of loop momenta. This property (which also persists at two loop level), extensively exploited  
in \cite{N46Sugra, Bern:2012gh}, implies that it is possible to construct four-point amplitudes in $\cN \ge 4$  
supergravities directly in terms of integrated gauge theory amplitudes. 
In contrast, the kinematic numerator amplitudes found in the previous section contain nontrivial loop 
momentum dependence.
Thus, they may be used to construct supergravity amplitudes at the integrand level.  We have checked that indeed the four-graviton amplitudes in the theories listed on rows
six, eight and thirteen of Table~\ref{taborb_moduli} are correctly reproduced. To reproduce amplitudes in the other theories  though the double-copy construction one only requires an  integrand representation of the numerator factors in pure YM theory coupled with some number of scalar fields.

While the parity-odd terms  present in the numerator factors (\ref{chiralmultiplet}) trivially integrate to zero in the gauge 
theory amplitudes they make nontrivial contributions to the corresponding supergravity amplitude; in particular, they are required for the UV finiteness of the four-graviton amplitude (they also produce parity-odd terms which integrate to zero). 
Introducing a flag $\flag $ to highlight the importance these contributions (which should be taken to 1 for the actual amplitude), we present  the result of the integral reduction of the four-graviton amplitude in the ${\cal N}=2$ 
supergravity coupled to one hypermultiplet realized as a double-copy of the $\NeqOne$ sYM theory given in the color-kinematics-satisfying representation above:
\bea
{\cal M}_{4, \NeqOne\times \NeqOne}^{(1)} \!\!\!&=&\!\!\!
{\cal M}_{4, \NeqEight}^{(1)}
+i\left(\frac{\kappa}{2}\right)^4M_\text{tree}
\Big[\,6\,\frac{tu}{s^2} (s^2-\frac{3(1+\flag)}{4}  {t u}) \,I_4^{D=6-2\epsilon}(t, u)
\\
                                          &&\!\!\!\!
                                              +\frac{9}{4}\,(1+\flag ) \frac{tu}{s^2}(t-u)(I_2(u)-I_2(t))              
                                              +\frac{9}{2}\frac{1+(3-D)\flag }{D-2}(I_2(u)+I_2(t))  \Big]\Big|_{\flag \rightarrow 1}\ .                               
\nonumber
\eea
%
%
%
%
%
%
This expressions reproduces eq.~(\ref{ampgrav}) with coefficients given by row~eleven of Table~\ref{tabAmpgrav}
which was derived though the direct evaluation of supergravity cuts. 
Through a similar strategy we can easily construct the one-loop four-graviton amplitude for the supergravity theory 
on row~thirteen of Table~\ref{taborb_moduli}. Keeping the flag tracking the contribution of the terms that vanish upon integration in the sYM amplitude, it reads
\bea
{\cal M}_{4, \NeqTwo\times \NeqOne}^{(1)} \!\!\!&=&\!\!\!
{\cal M}_{4, \NeqEight}^{(1)}
+i\left(\frac{\kappa}{2}\right)^4M_\text{tree}
\Big[\,\frac{tu}{s^2} (5s^2-3(1+\flag) {t u}) \,I_4^{D=6-2\epsilon}(t, u)
\\
                                          &&\!\!\!\!
                                              +\frac{3}{2}\,(1+\flag )\frac{tu}{s^2}(t-u)(I_2(u)-I_2(t))                                               
                                              +3\frac{1+(3-D)\flag }{D-2}(I_2(u)+I_2(t)) \Big]\Big|_{\flag \rightarrow 1}\ ,
\nonumber                                               
\eea
%
%
%
%
%
%
and, as expected, reproduces eq.~(\ref{ampgrav}) with coefficients given by row~six of Table~\ref{tabAmpgrav}.

The third four-graviton amplitude that can be constructed from the color/kinematics-satisfying sYM amplitudes 
we have constructed is in the ${\cal N}=4$ supergravity coupled to two vector multiplets realized as a double-copy 
of the $\NeqTwo$ sYM theory. Reducing the result of the double-copy construction to the standard integral basis 
yields
\bea
{\cal M}_{4, \NeqTwo\times \NeqTwo}^{(1)} \!\!\!&=&\!\!\!
{\cal M}_{4, \NeqEight}^{(1)}
+i\left(\frac{\kappa}{2}\right)^4
M_\text{tree}
\Big[\,4\,\frac{tu}{s^2}(s^2-\frac{1+\flag }{2} t u) \,I_4^{D=6-2\epsilon}(t, u)
\\
                                          &&\!\!\!\!
                                              + (1+\flag )\frac{tu}{s^2}(t-u)(I_2(u)-I_2(t))                                               
                                              +2\frac{1+(3-D)\flag }{D-2}(I_2(u)+I_2(t)) \Big] \Big|_{\flag \rightarrow 1} \ ;
\nonumber                                               
\eea
%
%
%
%
%
%
it indeed reproduces eq.~(\ref{Neq4w2vectors}) which was derived though the direct evaluation of supergravity 
cuts. It also reproduces the result of the double-copy construction applied to $\NeqFour$ sYM theory and pure 
YM theory coupled to two scalar fields \cite{N46Sugra, Bern:2012gh}.

We have also evaluated a number of four-dimensional cuts of higher-loop and higher-point amplitudes in the
two presentations of $\NeqFour$ supergravity coupled with two vector multiplets and found that they are the same.
This suggests that the scattering amplitudes of 
${\cal N}=2|_{{\bf Z}_2}\times {\cal N}=2|_{{\bf Z}_2}$ and ${\cal N}=4\times {\cal N}=0|_{{\bf Z}_2\times {\bf Z}_2}$ 
theories are the same to all loop orders and to all multiplicities, at least at the
integrand level.

\newpage

\renewcommand{\theequation}{5.\arabic{equation}}
\setcounter{equation}{0}

\section{Discussion}

In this paper we discussed orbifold supergravity theories with $\NeqEight$ supergravity as parent 
that preserve at least minimal amount of supersymmetry; 
similarly to non-planar orbifold gauge theories (and contrary to their planar limit), loop amplitudes 
are not inherited from the parent and an explicit calculation is necessary.
Realizing $\NeqEight$ supergravity tree-level amplitudes through the KLT relations, we identified {\em all} 
different supergravity theories that can be obtained with an orbifold group that acts independently on 
the two $\NeqFour$ sYM amplitude factors. 
Using generalized unitarity in four dimensions we have computed 
large classes (up to relabeling and use of supersymmetry Ward identities)  of four-point one-loop 
amplitudes of all such supergravity theories. Before reduction to the usual basis of 
box, triangle and bubble integrals  the amplitudes are sums of tensor integrals whose rank depends on 
the amount of supersymmetry. After reduction to the integral basis all surviving triangle integrals can be 
organized in terms of the six-dimensional box integral.
Both before and after integral reduction the amplitudes exhibit a mild non-locality involving inverse 
powers of Mandelstam invariants; similar features have been previously observed in sYM theories 
with less-than-maximal supersymmetry.

Our results show that the twin theories with $\cN=2$ and $\cN=4$, sharing the same  bosonic field content and  scalar manifold, 
have different four-point one-loop scattering amplitudes. We have also observed that all four-point one-loop 
amplitudes of the two different realizations of the $\NeqFour$ supergravity with two vector multiplets are 
identical and argued that this equality persists to all loop orders and to all multiplicities as well as away 
from four dimensions. Thus, the two orbifolds, ${\cal N}=2|_{{\bf Z}_2}\times {\cal N}=2|_{{\bf Z}_2}$ 
and ${\cal N}=4\times {\cal N}=0|_{{\bf Z}_2\times {\bf Z}_2}$, appear to lead to the same theory. This $D$-dimensional equivalence 
is reminiscent of, if perhaps not quite as dramatic as, the  equivalence of the double-copy construction of ${\cal N}=16$ supergravity in three dimensions
using either two copies of ${\cal N}=8$ sYM theory arranged on color-kinematic satisfying cubic graphs or  two copies of the BLG theory arranged on color-kinematic satisfying quartic  graphs~\cite{CSm}.

We also analyzed examples in which the orbifold group action is not factorized. Among them are 
pure supergravity theories with ${\cal N} = 3$, ${\cal N} = 2$ and ${\cal N} = 1$ supersymmetry. 
It would be interesting to analyze them at higher loops and thus probe their UV properties, in particular 
at three loops, where the first divergence is expected to appear. More general, it would also be interesting 
to classify all supergravity theories that may be obtained from $\NeqEight$ supergravity though 
non-factorized orbifold group action and explore their properties.

The calculation of amplitudes in all theories with a factorized orbifold group action can also be approached 
though the expected double-copy property of supergravity amplitudes. While for one of the two gauge theories
only an integrand representation is sufficient, for the other one a representation satisfying color/kinematics 
duality is required. We have constructed a representation (valid in four dimensions) of the one-loop four-gluon 
amplitude of pure ${\cal N}=2$ and ${\cal N}=1$ sYM theories. 
Due to the special relations between numerator factors imposed by the kinematic Jacobi relations, the 
size of the \Ansatz~  needed for such a construction is substantially smaller than that used for the direct 
unitarity-based construction.
We have found that it is not possible to satisfy both the color/kinematics duality constraints 
and the cut conditions with a local choice of numerator factors for all contributing graphs and a certain amount 
of nonlocality in the form of inverse powers of Mandelstam invariants is required. This suggests that, in some sense, 
generic amplitudes will only have nonlocal representations.  This is exactly the case for multi-loop maximally supersymmetric five-point color-kinematic representations~\cite{loop5ptBCJ} as well as certain tree-level representations~\cite{virtuousTrees}. 
Using these representations we have recovered through the double-copy construction the results of the direct 
unitarity-based calculation. It should be stressed that $D$-dimensional representations of the same sYM 
amplitudes are required to construct supergravity amplitudes whose integrand vanishes identically in four dimensions.

It should be possible and interesting to construct  multi-loop and 
higher-multiplicity amplitudes obeying color/kinematics duality for $\NeqTwo$ and $\NeqOne$ sYM theories.  The structure of the duality satisfying representations presented here is  very suggestive that the gluonic $\gamma$ and $\beta$ functions of ref.~\cite{loop5ptBCJ} should encode the gluonic supersymmetric color-kinematic satisfying representations, albeit with potentially higher powers of the Gram determinant in the denominator to allow for the required additional 
loop-momentum dependence. 
Such constructions would further probe the need for nonlocal kinematic numerator factors as well as the properties of 
nonlocal integrands for scattering amplitudes. 

In any case, it is clear that powerful methods and structures have now brought the ability to directly probe the $S$-matrix of intriguingly more intricate (and potentially phenomenologically relevant) supersymmetric gauge and gravity theories -- not only integrated, but at the integrand level as well.  Indeed, integrand level considerations are an incredibly powerful perspective for exposing invariant unifying structure that remains true independent of regularization schemes. This type of data will undoubtedly prove useful in the addressing of many open theoretic questions such as the ultimate role of non-linear dualities, shift symmetries, and unifications.


\section*{Acknowledgments}
\vskip -.3 cm 
We thank Zvi Bern, Scott Davies, Renata Kallosh, Josh Nohle, Eduardo Serna and Arkady Tseytlin for useful discussions.  
This research was supported by the US National Science Foundation under grants PHY-1213183, PHY-08-55356 and PHY-0756174.
JJMC also gratefully acknowledges the support of the Stanford Institute for Theoretical Physics, and a grant from the John Templeton Foundation.
Any opinions, findings and conclusions or recommendations expressed in this
article are those 
of the authors and do not necessarily reflect the views of the NSF or of the John Templeton Foundation.


\newpage

\appendix

\def\theequation{A.\arabic{equation}}
\setcounter{equation}{0}

\section{Integral Reduction\label{appPV}}

In this section we list the integral reductions used to obtain the final expressions for the amplitudes
listed in the main body of the paper. The external momenta $q_1, q_2$ are assumed to be strictly massless and four-dimensional. 
We have the following identities involving scalar triangles with numerator factors,
\bea I_3\big( l^\mu ; q_1 , q_1 +q_2 \big) &= & {q_2^\mu -q_1^\mu \over 2 q_1 \cdot q_2} I_2 (q_1+q_2) - q^\mu_1 I_3(q_1,q_1+q_2) \ ,\no \\
I_3\big( l^\mu l^\nu ; q_1 , q_1 +q_2 \big) &= & {\eta^{\mu \nu}\over 2d -4} I_2(q_1+q_2) + {q_1^\mu q_1^\nu} 
\Big( {3 \over 4} {I_2 (q_1+q_2)\over q_1 \cdot q_2} +   I_3(q_1,q_1+q_2) \Big)-  \no \\ 
&&  {q_1^{(\mu} q_2^{\nu)} d \over 4  (d-2)} {I_2(q_1+q_2)\over q_1 \cdot q_2} - {q_2^\mu q_2^\nu \over 4 q_1 \cdot q_2} I_2(q_1+q_2) \ ,\no \\
I_3\big( l^\mu l^\nu l^\rho; q_1 , q_1 +q_2 \big) &= &- {\eta^{(\mu \nu} q_1^{\rho)} d  \over 4 (d-1)(d-2)} I_2(q_1+q_2)
- {\eta^{(\mu \nu} q_2^{\rho)} \over 4 (d-1)} I_2(q_1+q_2) - \no \\
&& {q_1^\mu q_1^\nu q_1^\rho} \Big({7d-6 \over 8 (d-1)}{ I_2(q_1+q_2) \over q_1  \cdot q_2} 
+ I_3(q_1,q_1+q_2) \Big) + \no \\ 
&& {q_1^{(\mu}q_1^\nu q_2^{\rho)} d (d+2) \over 8  (d-1)(d-2)} {I_2 (q_1+q_2)\over q_1 \cdot q_2} + \no \\
 && {q_1^{(\mu}q_2^\nu q_2^{\rho)} (d+2)  \over 8 ( d-1)} 
 {I_2 (q_1+q_2)\over q_1 \cdot q_2} + {q_2^{\mu}q_2^\nu q_2^{\rho} d  \over 8 (d-1)} { I_2 (q_1+q_2) \over q_1 \cdot q_2} \ ,\no \\ 
I_3\big( l^\mu l^\nu l^\rho l^\sigma; q_1 , q_1 +q_2 \big) &= &
-{\eta^{(\mu \nu} \eta^{\rho \sigma)}\over 4 d (d-1) } (q_1\cdot q_2) I_2(q_1+q_2)+ {\eta^{(\mu \nu} q_1^{\rho} q_1^{\sigma)} (d+2) \over 8 (d-1) (d-2)} I_2(q_1+q_2) +\no \\
&&  {\eta^{(\mu \nu} q_1^{\rho} q_2^{\sigma)} (d+2) \over 8 d (d-1)} I_2(q_1+q_2)
 + {\eta^{(\mu \nu} q_2^{\rho} q_2^{\sigma)} \over 8 (d-1)} I_2(q_1+q_2)+ \no \\
 &&  {q_1^\mu q_1^\nu q_1^\rho q_1^\sigma} \Big({5 (3d-2) \over 16 (d-1)}{ I_2(q_1+q_2) \over q_1  \cdot q_2} 
+ I_3(q_1,q_1+q_2) \Big) - \no \\ 
&&  {q_1^{(\mu}q_1^\nu q_1^{\rho} q_2^{\sigma)} (d+2)(d+4) \over 16  (d-1)(d-2)} { I_2 (q_1+q_2) \over q_1 \cdot q_2}- \no \\
&& {q_1^{(\mu}q_1^\nu q_2^{\rho} q_2^{\sigma)} (d+2)(d+4) \over 16  d (d-1)} {I_2 (q_1+q_2) \over q_1 \cdot q_2}-\no \\
&&  {q_1^{(\mu}q_2^\nu q_2^{\rho} q_2^{\sigma)} (d+4)  \over 16 ( d-1)} {I_2 (q_1+q_2) \over q_1 \cdot q_2}- 
  {q_2^{\mu}q_2^\nu q_2^{\rho} q_2^{\sigma} (d+2)  \over 16 ( d-1)} { I_2 (q_1+q_2)\over q_1 \cdot q_2} \ . \no \\ \eea
Note that to obtain the expressions listed in the main body of the paper it is sufficient to apply the
Passarino-Veltman reduction on the triangle integrals with up to four loop momenta in the numerator, 
as boxes do not have numerator factors and bubbles do not explicitly appear in the \Ansatz. 


\def\theequation{B.\arabic{equation}}
\setcounter{equation}{0}
\section{ Jacobi relations  \label{AppJacobi}}

We include here the Jacobi relations between box, triangle and bubble graph kinematic numerator factors.
\begin{eqnarray}
N_ 1(k_ 1,k_ 2,k_ 3,k_ 2+k_ 3+q)-N_ 3(k_ 1,k_ 2,k_ 3,k_ 3+q)-N_ 4(k_ 1,k_ 2,k_ 3,-q)&=&0
\cr
N_ 1(k_ 1,k_ 2,k_ 3,-k_1-q)-N_ 2(k_ 1,k_ 2,k_ 3, q)-N_ 5(k_ 1,k_ 2,k_ 3,q)&=&0
\cr
N_ 1(k_ 1,k_ 2,k_ 3,-q)-N_ 3(k_ 1,k_ 2,k_ 3,q-k_ 1)-N_ 6(k_ 1,k_ 2,k_ 3, q)&=&0
\cr
N_ 1(k_ 1,k_ 2,k_ 3,k_ 2+k_ 3+q)-N_ 2(k_ 1,k_ 2,k_ 3,-k_1+q)-N_ 7(k_ 1,k_ 2,k_ 3,-q)&=&0
\cr
N_ 3(k_ 1,k_ 2,k_ 3,-q)-N_ 2(k_ 1,k_ 2,k_ 3,q-k_ 1)+N_ 9(k_ 1,k_ 2,k_ 3,q - k_ 1)&=&0
\cr
N_ 3(k_ 1,k_ 2,k_ 3,k_ 3+q)-N_ 2(k_ 1,k_ 2,k_ 3,q)-N_ 8(k_ 1,k_ 2,k_ 3,-k_1-k_ 2-k_ 3-q)&=&0
\cr
&& \label{fullJacobi}
\\
N_ 4(k_ 1,k_ 2,k_ 3,k_ 3+q) + N_ 4(k_ 1,k_ 2,k_ 3,k_ 1+k_ 2+k_ 3-q) -N_{10}(k_ 1,k_ 2,k_ 3,-q)&=&0
\cr
N_ 5(k_ 1,k_ 2,k_ 3,-k_1-k_ 2-k_ 3+q) + N_ 5(k_ 1,k_ 2,k_ 3,-k_1-q) +N_{11}(k_ 1,k_ 2,k_ 3,-q)&=&0
\cr
N_ 6(k_ 1,k_ 2,k_ 3,k_ 1+q) + N_ 6(k_ 1,k_ 2,k_ 3,-k_2-q) -N_{10}(k_ 1,k_ 2,k_ 3,-k_1-k_ 2-q)&=&0
\cr
-N_8(k_ 1,k_ 2,k_ 3,k_ 4+q) - N_ 8(k_ 1,k_ 2,k_ 3,-k_2-q) -N_{12}(k_ 1,k_ 2,k_ 3,-k_1-k_ 3-q)&=&0
\cr
-N_7(k_ 1,k_ 2,k_ 3,-q) + N_ 7(k_ 1,k_ 2,k_ 3,-k_3-q) -N_{11}(k_ 1,k_ 2,k_ 3,q)&=&0
\cr
N_ 9(k_ 1,k_ 2,k_ 3,k_ 1+q) + N_ 9(k_ 1,k_ 2,k_ 3,-k_3-q) +N_{12}(k_ 1,k_ 2,k_ 3,q)&=&0
\nonumber
\end{eqnarray}
These relations imply that the three box integrals may be chosen as master graphs at one-loop in 
any gauge theory.

Further Jacobi relations link triangle and bubble graphs to tadpole graphs. Requiring that the latter 
vanish leads to:
\bea
N_4(k_1,k_2,k_3,q)+N_4(k_1,k_2,k_3,k_1+k_2+k_3-q)&=&0
\cr
N_4(k_1,k_2,k_3,q)+N_4(k_1,k_2,k_3,k_3-q)&=&0
\cr
N_5(k_1,k_2,k_3,q)+N_5(k_1,k_2,k_3,-k_1-q)&=&0
\cr
N_5(k_1,k_2,k_3,q)+N_5(k_1,k_2,k_3,-k_1-k_2-k_3-q)&=&0
\cr
N_6(k_1,k_2,k_3,q)+N_6(k_1,k_2,k_3,-k_2-q)&=&0
\cr
N_6(k_1,k_2,k_3,q)+N_6(k_1,k_2,k_3,k_1-q)&=&0
\cr
N_7(k_1,k_2,k_3,q)+N_7(k_1,k_2,k_3,-k_3-q)&=&0
\cr
N_7(k_1,k_2,k_3,q)+N_7(k_1,k_2,k_3,k_2-q)&=&0
\cr
N_8(k_1,k_2,k_3,q)+N_8(k_1,k_2,k_3,-k_2-q)&=&0
\cr
N_8(k_1,k_2,k_3,q)+N_8(k_1,k_2,k_3,-k_1-k_2-k_3-q)&=&0
\cr
N_9(k_1,k_2,k_3,q)+N_9(k_1,k_2,k_3,-k_3-q)&=&0
\cr
N_9(k_1,k_2,k_3,q)+N_9(k_1,k_2,k_3,k_1-q)&=&0
\nonumber
\eea
These relations constrain the parameters of the \Ansatz~ for the kinematic numerators for the 
master graphs though their relation to the triangle numerator factors implied by the Jacobi relations.

\end{document}